\documentclass[12pt]{article}
\usepackage[dvips]{graphicx}
\usepackage{amssymb}
\usepackage{amsmath}
\usepackage{here}
\usepackage{color}
\usepackage{subfigure}
\usepackage{cite}

\newcommand{\be}{\begin{equation}}   
\newcommand{\ee}{\end{equation}}

\begin{document}

\begin{titlepage}

\begin{flushright}
ICRR-Report-???\\
UT-??
\end{flushright}

\vskip 2.0cm

\begin{center}

{\large \bf Cosmological and astrophysical constraints on \\superconducting cosmic strings}

\vskip 1.2cm

Koichi Miyamoto$^a$ and Kazunori Nakayama$^{b,c}$

\vskip 0.4cm

{\it $^a$Institute for Cosmic Ray Research,
University of Tokyo, Kashiwa 277-8582, Japan}\\
{\it $^b$Department of Physics, University of Tokyo, Bunkyo-ku, Tokyo 113-0033, Japan}\\
{\it $^c$Kavli Institute for the Physics and Mathematics of the Universe, University of Tokyo,
Kashiwa 277-8583, Japan}

\vskip 1.2cm

\date{\today}

\begin{abstract}
We investigate the cosmological and astrophysical constraints on superconducting cosmic strings (SCSs).
SCS loops emit strong bursts of electromagnetic waves, which might affect various cosmological and astrophysical observations.
We take into account the effect on the CMB anisotropy, CMB blackbody spectrum, BBN,
observational implications on radio wave burst and X-ray or $\gamma$-ray events, and
stochastic gravitational wave background measured by pulsar timing experiments.
We then derive constraints on the parameters of SCS from current observations 
and estimate prospects for detecting SCS signatures in on-going observations. 
As a result, we find that these constraints exclude 
broad parameter regions, and also that on-going radio wave observations can probe large parameter space.
\end{abstract}

\end{center}
\end{titlepage}

\section{Introduction}

Cosmic strings~\cite{Vilenkin} are one-dimensional massive objects,
which may appear as topological defects at the spontaneous symmetry breaking (SSB) in the early Universe~\cite{Kibble:1976sj}.
As a result of nonlinear dynamics, they form highly complicated string networks,
consisting of infinite strings, which stretch across the Hubble horizon, and closed loops.
They leave cosmological and astrophysical signatures in many ways.
As an important probe of both the early Universe and particle physics beyond the Standard Model,
they are investigated extensively.

There is a possibility that a cosmic string is superconducting, which carry
an electric current flowing without resistance~\cite{Witten:1984eb}.
Such a cosmic string is called a superconducting cosmic string (SCS).
Whether a cosmic string is superconducting or not depends on the way of SSB in the background particle physics model and such models can be constructed in Grand Unified Theory in natural ways~\cite{Witten:1984eb}.
SCSs are essentially characterized by two parameters.
One is the tension $\mu$, the mass per unit length, which is often written in the dimensionless form $G\mu$, 
where $G$ is the Newton constant, and the other is the magnitude of the current $I$.

A large electric current in a core of a SCS can leave unique signals of SCSs in many ways.
In their nonlinear dynamics, current-carrying string loops emit strong bursts of electromagnetic waves (EMWs)~\cite{Vilenkin:1986zz,Spergel:1986uu,Copeland:1987yv,BlancoPillado:2000xy}, 
in addition to gravitational wave (GW) bursts.
Such EMWs may lead to various types of observable effects. For example,
it leads to the distortion of the energy spectrum of the cosmic microwave background (CMB)~\cite{Sanchez:1988ek,Sanchez:1990kj,Tashiro:2012nb},
the early reionization of the Universe and its effect on the CMB power spectrum~\cite{Tashiro:2012nv}, 
gamma ray bursts (GRBs)~\cite{Babul:1986wd,Paczynski1988,Berezinsky:2001cp,Cheng:2010ae},
\footnote{
However, there are objections against the idea that SCSs can be engines of GRBs~\cite{Wang:2011zc}.
}
radio transients~\cite{Vachaspati:2008su,Cai:2011bi,Cai:2012zd} and so on.
Besides, massive particles might be created from SCSs and their decay products may be detected as ultra high energy cosmic rays\cite{Hill:1986mn,Berezinsky:2009xf}.

While these observable signatures of SCSs have been studied,
there are not so many papers which focused on the constraints on the parameters characterizing SCSs, 
$G\mu$ and $I$, from current observations or the parameter regions which can be probed in future experiments.\footnote{
In Refs.~\cite{Tashiro:2012nb,Tashiro:2012nv} constraints from the CMB observation are studied. 
}
In this paper, we aim to derive constraints or future prospects not only integrating signals studied previously 
but also considering novel observable effects of SCSs.

We consider following types of signals of SCSs.
The first one is the change of the ionization history of the Universe and its effect on the power spectrum of the CMB anisotropy.
While Ref.~\cite{Tashiro:2012nv} focused on the effects of SCSs on reionization,
EMWs injected at any time after the recombination around the redshift $z_{\rm rec}\simeq 1090$
can ionize neutral hydrogens and alter the evolution of the ionization fraction.
In fact, in the context of decaying or annihilating dark matter, the effects on the ionization history of the energy injection in the whole epoch after recombination have been studied~\cite{Chen:2003gz,Zhang:2007zzh,Padmanabhan:2005es,Kanzaki:2009hf,Hisano:2011dc}.
Because of the high accuracy of CMB measurements, the energy injection at any redshift smaller than 
$z_{\rm rec}$ is constrained severely.
We will see that for SCSs, the most stringent constraint from the CMB comes from the energy injection around the recombination epoch.

The second one is the CMB spectral distortion.
The energy injection from SCS loops into the baryon-photon fluid between $z\simeq 2\times 10^6$ and $z\simeq 5\times 10^4$ induces the $\mu$-distortion of the energy spectrum of CMB photons from a blackbody spectrum, and that between $z\simeq 5\times 10^4$ and $z\simeq 1090$ induces the $y$-distortion\cite{SunyaevZeldovich,Danese1,Danese2,Burigana1,Burigana2,Hu:1992dc}.
In addition to EMW emission, which was considered in Ref.~\cite{Tashiro:2012nb}, 
the energy losen by dissipation into the plasma is also causes this effect.

The third one is the effect on Big Bang Nucleosynthes (BBN).
Since the observed abundance of light elements are consistent with the prediction of BBN,
the energy injection in the BBN epoch, which destroy nuclei and alter their abundance, is severely constrained~\cite{Kawasaki:2004yh,Jedamzik:2004er}.
We can derive constraints on SCSs from BBN so that
EMWs radiated by SCS loops do not photodissociate these light elements too much.

The fourth one is direct observations of EMWs from SCS loops.
EMWs emitted after recombination can be directly observed by current telescopes in various frequency bands,
unless they are absorbed in their propagation.
In this paper, we consider the possibility that photons from SCS loops are detected as radio waves 
or high energy photons such as X-rays or $\gamma$-rays.
Since radio wave bursts from SCS loops have very short duration, $\Delta \lesssim 10^{-2} {\rm s}$ as mentioned below,
they are observed as radio transients by telescopes such as Parkes~\cite{Parkes}.
In fact, there is an argument that an extragalactic radio wave burst has been discovered in the data of Parkes~\cite{Lorimer:2007qn}.
High energy photons such as X-rays and $\gamma$-rays have been also observed extensively in many experiments 
including the Fermi satellite~\cite{Fermi}. 
As mentioned below, bursts of X-rays or $\gamma$-rays originating from SCS loops have duration much shorter than the time resolution of the instruments, hence many photons fly in the detector simultaneously in the events of SCSs.
This makes SCSs signals distinguishable from those of the background or astrophysical events such as GRBs, which last much longer.

Finally, we consider the limit on the stochastic GW background emitted by SCSs from pulsar timing experiments.
For SCSs, since the current $I$ affects the number density of loops and the ratio of energy emitted as GWs to that as EMWs, 
the pulsar timing limit on the tension $G\mu$ depends on $I$ in the case that EMW emission is more efficient than GW emission.

Integrating the above cosmological and astrophysical signatures of SCSs, we find the excluded regions in $I$-$G\mu$ plane and the regions which can be probed by on-going observations.
We will find that the CMB anisotropy constraint gives the most stringent constraints, while
the CMB distortion constraint, the BBN constraint and the constraint from the past data of radio burst search at Parkes also exclude some parameter spaces
which the CMB anisotropy observation is not sensitive to.
Out of the excluded regions, there are large areas in the parameter plane which can be probed by the HTRU survey~\cite{Keith:2010kk}, the on-going observation at Parkes Observatory, 
while the parameter area in which Fermi satellite can detect signals of SCSs at a considerable rate is limited.

Concerning loop size, we consider the two cases : the large-loop case and the small-loop case.
In the former case, the size of loops is about a tenth of the Hubble radius at the formation 
and they typically survive much longer time than one Hubble time.
In the latter case, on the other hand, loops are so small that they disappear in one Hubble time, 
radiating their energy as GWs and EMWs.
The actual loop size remains unclear because of the difficulty to trace the highly nonlinear dynamics of the string network, 
so we consider these two typical cases about loop size.
We will see that in the large-loop case the excluded parameter regions or regions which can be probed in the future are larger than in the small-loop cases, since large-loops can survive longer cosmic time and hence energy density of loops 
can be much larger if they were born in the radiation dominated era.
On the evolution of SCS loops, we take into account energy dissipation due to the SCS interaction with surrounding plasma, which is induced by the current on loops~\cite{Vilenkin}, 
in addition to emission of EMWs and GWs.
This effect was neglected in most of recent papers on SCS.

This paper is organized as follows.
In Sec.~2, we briefly review the EMW emission by SCS loops and their abundance.
In Sec.~3, we consider the constraints mentioned above and derive excluded regions in the parameter plane of SCSs.
Moreover, we derive the contour plot of the detection rate of SCS events in various types of on-going observations to estimate the prospect for SCS searches in the near future.
We summarize this paper in Sec.~4.

In this paper, cosmological parameters are set as below, according to the latest result of WMAP~\cite{Komatsu:2010fb}. 
Density parameter of dark energy : $\Omega_\Lambda=0.728$, density parameter of matter : $\Omega_m=0.272$, the current temperature of the CMB : $T_0=2.725{\rm K}$, 
the present Hubble constant : $H_0=70.4{\rm km}/{\rm s}/{\rm Mpc}$.
We define the present value of the scale factor as $a_0=1$.

\section{Some basics about SCS loops}

\subsection{EMW and GW emission from SCS loops}

We assume that SCSs have an equal, uniform and constant current $I$.
\footnote{
The current around a cusp region reaches the chiral limit\cite{Spergel:1986uu,Copeland:1987yv,Davis:1988jq} and after that it is constant, unless there is external effects such as current induction due to the external magnetic field, which we do not consider in this paper.
}
A SCS loop with current emits EMW bursts from cusps, highly relativistic regions which appears $\mathcal{O}(1)$ times per one oscillation period.\footnote{
EMW bursts from kinks are studied in Ref.~\cite{Cai:2012zd}.
We do not consider them in this paper because the energy emitted from a kink per unit time is much smaller than 
that from a cusp.
}
The energy of EMWs emitted from a loop with length $l$ into unit solid angle per frequency in a cusp event is given by~\cite{Cai:2011bi}
\be
\frac{d^2E}{df d\Omega}\sim I^2l^2.
\label{dE_dfdOme}
\ee
A solid angle of the EMW emission from a cusp depends on the EMW frequency. It is given by
\be
\Delta \Omega(f) = \pi \theta^2_m(f), \ \theta_m(f)\sim (lf)^{-1/3}.
\label{thetam}
\ee
Since $l$ is order of cosmological scale, $lf \gg 1$ is satisfied for frequencies we are interested in. 
The energy emitted from a cusp event per unit frequency is
\be
\frac{dE}{df}\sim I^2l^2\Delta\Omega(f)\sim \pi I^2l^{4/3}f^{-2/3}.
\label{dE_df}
\ee
Note that $\int df dE/df$ is divergent, hence there must be a cutoff frequency $f_c$.
A model independent cutoff frequency comes from the requirement that the emitted EMWs
do not significantly back-react to the SCS motion.
It is given by~\cite{Vilenkin:1986zz,Spergel:1986uu,BlancoPillado:2000xy}
\be
\omega_c=2\pi f_c\sim \mu^{3/2}I^{-3}l^{-1}
\sim 10^9\,{\rm GeV}\left( \frac{G\mu}{10^{-7}} \right)^{3/2}
\left( \frac{1\,{\rm GeV}}{I} \right)^3\left( \frac{10^{13}\,{\rm sec}}{l} \right).
\label{omegac}
\ee
Therefore, the total energy emitted in a cusp event is given by
\be
E_{\rm tot}\sim Il\sqrt{\mu}.
\ee
The main contribution to the EMW emission comes from the frequency of $f\sim f_c$.
Since cusps appear on a loop at a time interval $\sim l$, the energy emitted from a loop per unit time is
\be
P_{\rm EM}=\Gamma_{\rm EM} I\sqrt{\mu},
\ee
where $\Gamma_{\rm EM}$, which is set to be $10$ in this paper, is a numerical constant.
A SCS loop also emits GWs.
The energy of GWs emitted from a loop per unit time is
\be
P_{\rm GW}=\Gamma_{\rm GW} G\mu^2,
\ee
where $\Gamma_{\rm GW}$ is a numerical constant which is taken to be $50$ in this paper.
For $I>I_*$ where
\begin{equation}
	I_* \equiv \frac{\Gamma_{\rm GW}G\mu^{3/2}}{\Gamma_{\rm EW}} \simeq 6\times 10^4\,{\rm GeV}
	\left( \frac{G\mu}{10^{-10}} \right)^{3/2},
\end{equation}
a loop loses its energy mainly by the EMW emission, i.e. $P_{\rm EM}>P_{\rm GW}$.
For $I<I_*$, on the other hand, the GW emission is the dominant channel of energy release, i.e. $P_{\rm GW}>P_{\rm EM}$.
Hereafter, we call the former case as the EMW-dominated case and the latter as the GW-dominated case.

Note that, for most parameter region of interest, the cutoff frequency (\ref{omegac}) is much larger than the plasma frequency $\omega_p$,
\begin{equation}
	\omega_p = \sqrt{\frac{4\pi \alpha_e n_e}{m_e}} \simeq  2\times 10^{-23}\,{\rm GeV} ~ x_e(z)^{1/2}(1+z)^{3/2},
\end{equation}
where $n_e$ is the number density of electron, $m_e$ is the electron mass, and $x_e(z)$ denotes
the ionization fraction of the hydrogen at the redshift $z$.
Therefore, we do not need to consider the damping of EMWs in the cosmological plasma.

It should also be noticed that the SCSs feel friction from the surrounding plasma~\cite{Vilenkin} and dissipate their energy into the plasma.
The energy loss rate of a SCS loop with length $l$ due to this effect is given by
\be
P_{\rm dis} \sim F_{\rm dis} l, \  F_{\rm dis} \sim
\begin{cases}
\rho_e \nu  & ; \ R_s\lesssim \nu \\
\rho_e R_s  & ; \ R_s\gtrsim \nu
\end{cases}
, \label{Pdis}
\ee
where $F_{\rm dis}$ is the force acting on the SCS per unit length, $\rho_e$ is the energy density of the electron, $\nu$ is the viscosity of the plasma and
$R_s\sim I/\sqrt{\rho_e}$ is the radius of the cylindrical region where plasma particles cannot invade due to the magnetic field induced by the current on the loop.
Since $R_s/\nu$ is an increasing function of time, the upper expression of (\ref{Pdis}) is valid at the earlier time and the lower one is valid at the later time. 
See Appendix for details.

\subsection{Evolution of SCS loops}

Next, let us estimate the abundance of SCS loops.
As a consequence of repeating collisions and reconnections, the string network reach a scaling regime,
where the network of infinite strings can be thought of as a random walk with correlation length $\xi$ comparable to the Hubble radius. 
In order to maintain the scaling, infinite strings must emit their energies in the form of loops.
Assuming the length of loops which are formed at time $t$ is $\alpha t$, 
the formation rate of loops at time $t$ per unit volume is given by
\be
\frac{dn}{dt}\sim \frac{1}{\gamma(t)^2 \alpha t^4}.
\ee
Here we have assumed $\xi=\gamma(t) t$ and $\gamma(t)$ changes like a step function,
\be
\gamma(t)=
\begin{cases}
\gamma_r  & ;t<t_{\rm eq} \\
\gamma_m  & ;t>t_{\rm eq}
\end{cases}
,
\ee
where $t_{\rm eq}$ is the time at the matter-radiation equality and the subscript $r$($m$) represents the value in the radiation (matter)-dominated era.
We take $\gamma_r=0.27$ and $\gamma_m=0.64$ in the following~\cite{Martins:2000cs}.
The number density of loops decrease in proportion to $a(t)^{-3}$, where $a(t)$ is the scale factor.
Therefore, the number density of loops at the time $t$, which are formed at the time $t_i$ is given by
\be
\frac{dn}{dt_i}(t,t_i)\sim \frac{1}{\gamma(t_i)^2 \alpha t_i^4}\left(\frac{a(t_i)}{a(t)}\right)^3.
\label{dn_dti}
\ee

After the formation, a loop shrinks radiating EMWs and GWs and dissipating its energy by the friction.
The rate of shrinkage is given by
\be
\dot{l}=-\Gamma_{\rm eff}-\frac{F_{\rm dis}}{\mu}l,
\ee
where
\be
\Gamma_{\rm eff}=\Gamma_{\rm GW}G\mu+\Gamma_{\rm EM}\frac{I}{\sqrt{\mu}},
\ee
represents the efficiency of GW and EMW radiation and the dot denotes time derivative.
If dissipation can be neglected, i.e., $P_{\rm EM}+P_{\rm GW}>P_{\rm dis}$, the length of a loop decreases proportionally to time as
\be
l(t,t_i)=\alpha t_i -\Gamma_{\rm eff}(t-t_i),
\label{l_t}
\ee
where $t_i$ is the time when the loop is formed.
The lifetime of a loop is given by
\be
\tau=\frac{\alpha}{\Gamma_{\rm eff}} t_i
\ee
and a loop with size $l=l_d(t)$, where
\be
l_d(t)=\Gamma_{\rm eff} t,
\ee
decays within a Hubble time emitting GWs and EMWs.
On the other hand, plasma dissipation forces a loop to shrink exponentially.
If
\be
({\tau}_{\rm dis}H)^{-1}>1,
\ee
where
\be
{\tau}_{\rm dis}^{-1}=\frac{F_{\rm dis}}{\mu}
\ee
and $H=\dot{a}/{a}$ is the Hubble parameter, a loop decays within a Hubble time dissipating almost all of its energy into the plasma.
This does not depend on the size of the loop, unless $l<l_d(t)$.
The condition $({\tau}_{\rm dis}H)^{-1}>1$ also means that plasma dissipation is the dominant energy loss mechanism, i.e., $P_{\rm dis}>P_{\rm GW}+P_{\rm EM}$ for loops with length $l>l_d(t)$.
Hereafter, we call the period of $({\tau}_{\rm dis}H)^{-1}>1$ the dissipation-dominated period.
Note that, as a function of time, $({\tau}_{\rm dis}H)^{-1}$ has a peak at the time when $T=m_e$ if $R_s<\nu$, while it is an increasing function in the radiation-dominated era and constant between the matter-radiation equality and the recombination if $R_s>\nu$.
Therefore, there can be two dissipation-dominated periods. 
Depending on $G\mu$ and $I$, either of them might not exist or they might be combined.

Despite many numerical and analytical studies on the dynamics of the string network, the magnitude of $\alpha$ remains unclear: it varies from $0.1$ to some power of $G\mu$.\footnote{
As for studies on the loop size, we refer to Refs.~\cite{Bennett:1987vf,Allen:1990tv,Vincent:1996rb,Vanchurin:2005pa,Ringeval:2005kr,Martins:2005es,Olum:2006ix,Siemens:2002dj,Polchinski:2006ee,Dubath:2007mf,Vanchurin:2007ee,BlancoPillado:2011dq}.
}
We then consider the two cases separately in the following analyses.
One is the large-loop case, where
\be
\alpha=0.1.
\ee
In this case, since $\alpha\gg\Gamma_{\rm eff}$, a loop is long-lived : 
it survives much longer than one Hubble time, as long as GW or EMW emission is the dominant mechanism of energy loss.
The other is the small-loop case, where
\be
\alpha=\Gamma_{\rm eff}.
\ee
In this case, a loop is short-lived : it disappears within one Hubble time after the formation,
independently of whether the dominant energy release channel is EMW emission, GW emission or plasma dissipation.
If loops are long-lived, the loop density is enhanced compared with the background energy density in the radiation-dominated era,
since the loop density is proportional to $a^{-3}$ while the radiation density is proportional to $a^{-4}$.
This makes the CMB anisotropy, the CMB distortion and the BBN constraints in the large-loop case severer than those in the small-loop case and
the detection rate of EMW bursts higher, as we will see below.
In both the large-loop case and the small-loop case, 
loops about to disappear is dominant in term of number and energy density over loops of different length.
Length of such loops at time $t$ is given by $l\sim\alpha t_i(t)$, where $t_i(t)$ is the time when they are formed.
$t_i(t)$ is given by
\be
t_i(t)=
\begin{cases}
t & ; \ ({\tau}_{\rm dis}(t)H)^{-1}>1 \\
\max \{ t_{\rm dis} , \frac{\Gamma_{\rm eff}}{\alpha}t \} & ; \   ({\tau}_{\rm dis}(t)H)^{-1}<1
\end{cases}
,
\ee
where $t_{\rm dis}$ is the latest time when $({\tau}_{\rm dis}(t)H)^{-1}$ exceeds 1 before $t$.
If such $t_{\rm dis}$ does not exist, $t_i(t)=\frac{\Gamma_{\rm eff}}{\alpha}t$ and $l\sim l_d(t)$.
We consider only loops about to disappear below.
The results in the following sections do not change much even if loops of all length are taken into account.
The number density of disappearing loops are given by
\be
n_d(t)\sim \frac{1}{\gamma(t_i(t))^2 \alpha t_i(t)^3}\left(\frac{a(t_i(t))}{a(t)}\right)^3.
\ee
The energy density of disappearing loops is $\rho_d(t)=n_d(t)\times \mu \alpha t_i(t)$. 
In the large-loop case, if plasma dissipation is negligible, we obtain
\be
\rho_d(t)\sim
\begin{cases}
\displaystyle
\frac{\mu}{\gamma_r^2 t^2}\left(\frac{\alpha}{\Gamma_{\rm eff}}\right)^{1/2} & ; t<t_{\rm eq} \\
\displaystyle
\frac{\mu t_{\rm eq}^{1/2}}{\gamma_r^2t^{5/2}}\left(\frac{\alpha}{\Gamma_{\rm eff}}\right)^{1/2} & ; t_{\rm eq}<t<\frac{\alpha}{\Gamma_{\rm eff}}t_{\rm eq} \\
\displaystyle
\frac{\mu}{\gamma_m^2 t^2} & ; t>\frac{\alpha}{\Gamma_{\rm eff}} t_{\rm eq}
\end{cases}
.
\ee
Note that $\mu/\gamma^2t^2$ is the energy density of infinite strings.
In the radiation-dominant era, the energy density of disappearing loops is enhanced by a factor $\left(\alpha /\Gamma_{\rm eff}\right)^{1/2}$ as mentioned above.
On the other hand, in the small-loop case or if $({\tau}_{\rm dis}H)^{-1}>1$, we obtain $\rho_d(t)\sim\mu/\gamma(t)^2t^2$. 
It means that the energy density of loops is equal to that of infinite strings since they disappear right after their formation.

\subsection{A model of SCS}

Here we shortly point out that SCS may naturally arise in a class of supersymmetric (SUSY) models.
We introduce two gauged U(1)'s, U(1)$_{X}$ and U(1)$_{Y}$.
Let us take the superpotential as
\begin{equation}
	W =  X(\kappa_\phi \phi\bar\phi + \kappa_\chi \chi\bar\chi  - \mu^2),
\end{equation}
where $X$ is a singlet chiral superfield, $\phi (\bar\phi)$ are charged under U(1)$_X$ gauge group
and $\chi (\bar\chi)$ are charged under U(1)$_Y$ gauge group,
and $\kappa_\phi$ and $\kappa_\chi$ are coupling constants taken to be real and positive.
After including the SUSY breaking effect, the scalar potential is given by
\begin{equation}
	V = |\kappa_\phi \phi\bar\phi + \kappa_\chi \chi\bar\chi  - \mu^2|^2 + 
	m_\phi^2(|\phi|^2 + |\bar\phi|^2) + m_\chi^2(|\chi|^2 + |\bar\chi|^2),
\end{equation}
where we have set $X=0$ and taken the soft masses for $\phi$ and $\bar\phi$ to be same for simplicity, 
and also similarly for $\chi$ and $\bar\chi$.\footnote{
	Strictly speaking, $X$ takes finite expectation value if one correctly takes into account the constant term
	in the superpotential. But it is much smaller than the scale $\mu$ and safely ignored in the present discussion.
}
We have also omitted the D-term potential, which enforce $|\phi|=|\bar\phi|$ and $|\chi|=|\bar\chi|$.
This is quite similar to the scalar potential that allows the SCS solution~\cite{Witten:1984eb}.
Let us see some more details.
As explicitly recognized in Refs.~\cite{Hindmarsh:2012wh,Nakayama:2012gh},
there is a moduli space along the direction satisfying $\kappa_\phi\phi\bar\phi+\kappa_\chi\chi\bar\chi = \mu^2$ without soft masses,
and this degeneracy is broken due to the effect of soft mass terms.
We find that the potential minimum is given by either the point P1 or P2 in the scalar field space :
\begin{equation}
\begin{split}
	{\rm P1} : \langle |\phi|\rangle = 0, ~~\langle |\chi|\rangle = \mu^2/\sqrt{\kappa_\chi} ~~{\rm for~~}m_\phi^2/\kappa_\phi > m_\chi^2/\kappa_\chi, \\
	{\rm P2} : \langle |\chi|\rangle =0,~~\langle |\phi|\rangle =  \mu^2/\sqrt{\kappa_\phi} ~~{\rm for~~}m_\phi^2/\kappa_\phi < m_\chi^2/\kappa_\chi.
\end{split}
\end{equation}

Let us consider the latter case : $m_\phi^2/\kappa_\phi < m_\chi^2/\kappa_\chi$ where the point P2 is the true minimum.
To see how SCS is formed, we take this model as a hybrid inflation sector with $X$ being identified with
the inflaton field~\cite{Copeland:1994vg}.
In this model, it is found that the tachyonic instability first develops on the direction of $\chi$ if $\kappa_\chi < \kappa_\phi$ :
$\chi$ takes a role of waterfall field in this case.
Therefore, if these conditions are satisfied, scalar fields first fall into P1 just after hybrid inflation and then
finally they find the true vacuum P2.
We identify U(1)$_X$ as, e.g., the B$-$L gauge group, which is spontaneously broken by the large VEV of $\phi$ at the present vacuum
and U(1)$_Y$ as the EM gauge group or others including it.
It is expected that this series of phase transitions produce SCS,
although soft mass scales are typically much smaller than $\mu$ and hence phenomenology of such thick
strings might be different from ordinary ones.\footnote{
	Thermal inflation would take place during the course of this final phase transition~\cite{Hindmarsh:2012wh}.
}

\section{Constraints on parameters of SCSs}

\subsection{CMB anisotropy constraint}

The anisotropy of the CMB has been observed in various experiments with high accuracy.
Because of high accuracy of CMB measurements, nonstandard energy injections which lead to the extra ionization 
are severely constrained over the whole epoch after recombination.
In the context of decaying or annihilating dark matter, the limit on energy injection is studied in 
Refs.~\cite{Chen:2003gz,Zhang:2007zzh,Padmanabhan:2005es,Kanzaki:2009hf,Hisano:2011dc}.
The energy of EMWs emitted from SCS loops per unit volume per Hubble time is given by
\be
\rho_{\rm inj}(t)\sim n_d(t)\Gamma_{\rm EM}I\sqrt{\mu}t\sim \rho_d(t)\times \frac{P_{\rm EM}}{P_{\rm tot}},
\ee
where $P_{\rm tot}=P_{\rm EM}+P_{\rm GW}+P_{\rm dis}$.
Note that $\rho_{\rm inj}(t)/\rho_{\rm tot}$ is a decreasing or constant function of $t$ in the matter-dominated era.
This means that the most stringent constraint comes from the energy injection around the recombination epoch.
From Ref.~\cite{Zhang:2007zzh}, 
the upper bound on the injected energy by SCS loops is inferred as 
$\rho_{\rm lim}=7.0\times 10^{-32} {\rm g}/{\rm cm^3}$ at $t=10^{13}\,{\rm s}$.
Then the condition 
\be
\rho_{\rm inj}(t=10^{13}{\rm s})<\rho_{\rm lim} \label{CMB}
\ee
 leads to constraint on $I-G\mu$ plane.

\begin{figure}[p]
\begin{minipage}{1\hsize}
\begin{center}
\subfigure[The large-loop case, $\alpha=0.1$. Under the black dashed line, $\dot{\tau}_{\rm dis}H^{-1}>1$ at the recombination.]{
\includegraphics[width=110mm]
{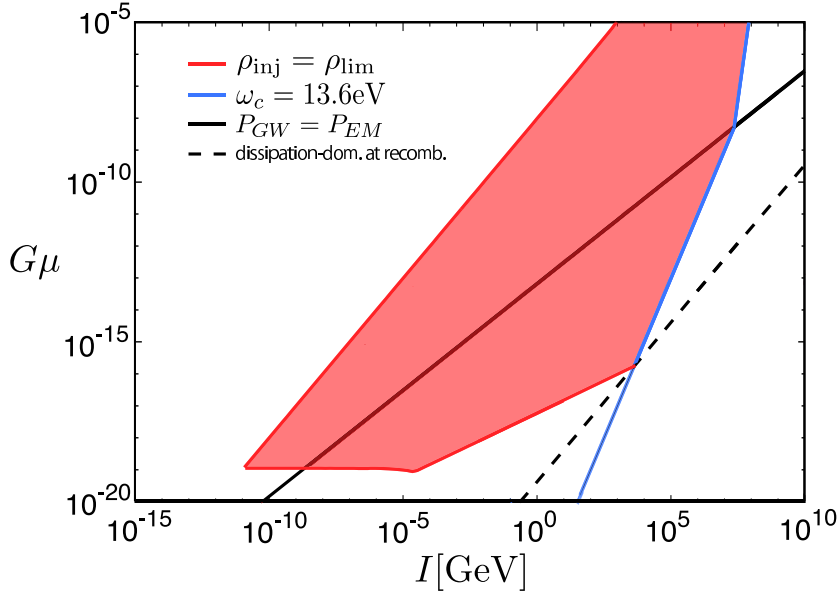}
}
\subfigure[The small-loop case, $\alpha=\Gamma_{\rm eff}$]{
\includegraphics[width=110mm]
{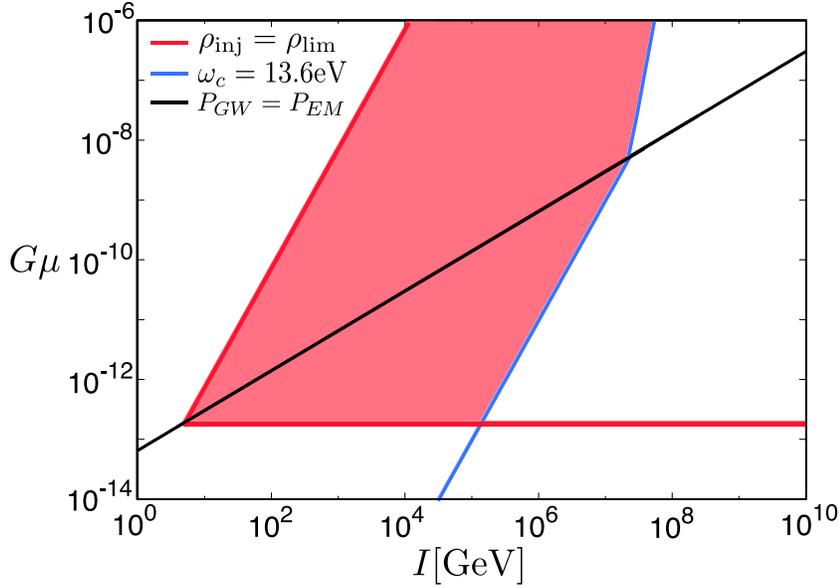}
}

\end{center}
\end{minipage}
\caption{Constraint from the CMB anisotropy observation in the large-loop case (top panel) and the small-loop case (bottom panel).
In each figure, between the two red lines the injected energy is larger than $\rho_{\rm lim}$. 
In the left side of the blue line the cutoff energy of emitted photons from SCS loops is larger than the ionization energy of hydrogen.
Red regions are excluded.
On the black line the energy radiated by EMW emission per unit time is equal to that by GW emission.}
\label{fig:CMBconstraint}
\end{figure}

Even though the injected energy exceeds $\rho_{\rm lim}$, 
there is no significant effects on the CMB anisotropy if the energy of photons emitted by SCS loops is too low to ionize hydrogens.
The cutoff energy of emitted photon, given by (\ref{omegac}), shows that photons emitted earlier have larger energy than those emitted later, since the loop size $l$ is smaller at earlier time.
This is the case if $\omega_c=\mu^{3/2}I^{-3}(l_d(t=10^{13}{\rm s}))^{-1}<13.6\,{\rm eV}$.
This condition is rewritten as
\be
G\mu <
\begin{cases}
3.3\times 10^{-11}\left(\frac{I}{10^7\,{\rm GeV}}\right)^6 & ;I<I_* \\
9.4\times 10^{-10}\left(\frac{I}{10^7\,{\rm GeV}}\right)^2 & ;I>I_*
\end{cases}
\label{omegacCMB}
\ee
in both the large-loop case and the small-loop case.

Parameter regions which violate both (\ref{CMB}) and (\ref{omegacCMB}) are excluded by CMB anisotropy observations.
The excluded regions in $I$-$G\mu$ plane in the two cases are shown in Fig.~\ref{fig:CMBconstraint}.
The shape of the excluded region is explained as follows.
In the GW-dominated case, $I<I_*$, the fraction of energy emitted as EMWs to that as GWs becomes smaller as $G\mu$ increases.
Thus a lower bound of $G\mu$ is obtained.
On the other hand, in the EMW-dominated case, the energy of disappearing loops is radiated almost as EMWs, 
and the energy of such EMWs becomes larger as $G\mu$ increases.
Therefore, $G\mu$ is bounded from above in the EMW-dominated case.
For a small value of $G\mu$, plasma dissipation must be considered.
Under the black dashed line in the large-loop case of Fig.~\ref{fig:CMBconstraint}, $({\tau}_{\rm dis}H)^{-1}>1$ at the recombination.
Therefore, energy injection by EMW emission around the recombination is strongly suppressed.
Even after it, energy injection as EMWs is suppressed since all loops are generated after the recombination and the enhancement of loop number density mentioned in Sec 2.1 does not occur.
However, the excluded parameter region is above this line, so the effect of plasma dissipation around the recombination does not affect the result.
On the other hand, around the bottom of the excluded region, where $G\mu\sim 10^{-19}$ and $10^{-11}{\rm GeV}\lesssim I\lesssim 10^{-5} {\rm GeV}$, dissipation plays an important role.
There is a dissipation-dominated period before the recombination.
This leads to suppression of energy injection around the recombination for parameter sets under the red horizontal line.
In the small-loop case, plasma dissipation dose not affect the constraint since loops at time $t$ are generated just before $t$ and dissipation effect is negligible after the recombination until the reionization.

The largeness of the excluded region in Fig.~\ref{fig:CMBconstraint} shows that CMB anisotropy observations strongly constrain SCSs in terms of $G\mu$ and $I$.
In fact, the CMB anisotropy constraint excludes the larger area in $I$-$G\mu$ than any other constraints considered below.
The excluded area in the large-loop case is much larger than that in the small-loop case, because of the aforementioned effect that the loop density is enhanced.

\subsection{CMB spectral distortion constraint}

The energy injection before recombination can induce deviation of the energy spectrum of the CMB photons from the blackbody spectrum~\cite{SunyaevZeldovich,Danese1,Danese2,Burigana1,Burigana2,Hu:1992dc}.
After $z\simeq 2\times 10^6$, the double-Compton scattering, which is the photon number violating process and necessary  to maintain chemical equilibrium between photons and electrons, becomes inefficient.
Therefore, photons injected after $z\simeq 2\times 10^6$ change the energy spectrum of photons so that it has the nonzero chemical potential $\mu$.
This type of the CMB spectral distortion is called $\mu$-distortion.
After $z\simeq 5\times 10^4$, the Compton scattering, which is necessary to re-distribute the injected energy, also becomes inefficient.
Photons injected after that induce the so-called $y$-distortion of the energy spectrum of the CMB photons, which can be characterized by the Compton $y$ parameter.
These types of CMB distortion are constrained by COBE FIRAS measurement; $|\mu|<9\times 10^{-5}, \ y<1.5\times 10^{-5}$~\cite{Mather:1993ij,Fixsen:1996nj}.
The future satellite, called PIXIE, might have much better sensitivity on these CMB distortion parameters; $|\mu|\sim 10^{-8}, \ y\sim2\times 10^{-9}$ at the $1\sigma$ level.

\begin{figure}[p]
\begin{minipage}{1\hsize}
\begin{center}
\subfigure[The large-loop case, $\alpha=0.1$. Under the dashed and dotted line, $({\tau}_{\rm dis}H)^{-1}>1$ over the epochs when $\mu$ and $y$ distortions are generated, respectively.]{
\includegraphics[width=110mm]
{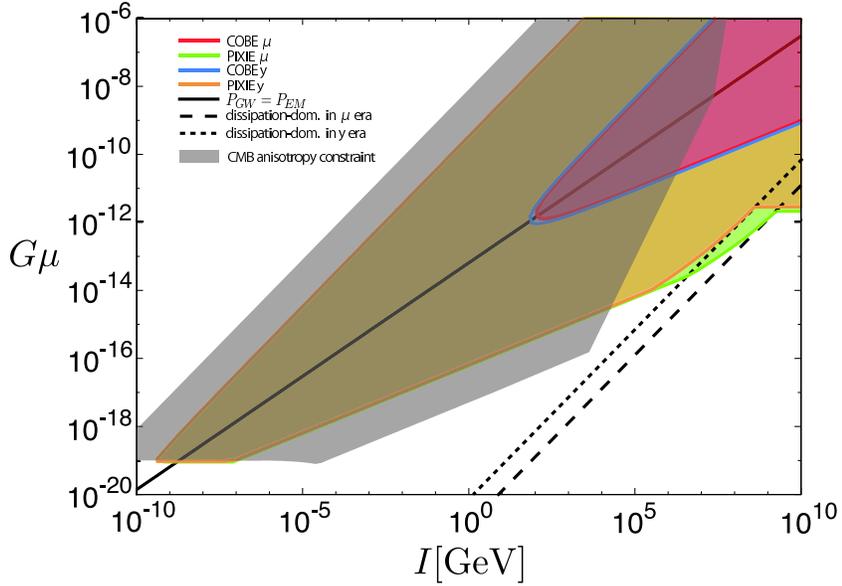}
\label{fig:CMBdistconstraintlarge}
}
\subfigure[The small-loop case, $\alpha=\Gamma_{\rm eff}$]{
\includegraphics[width=110mm]
{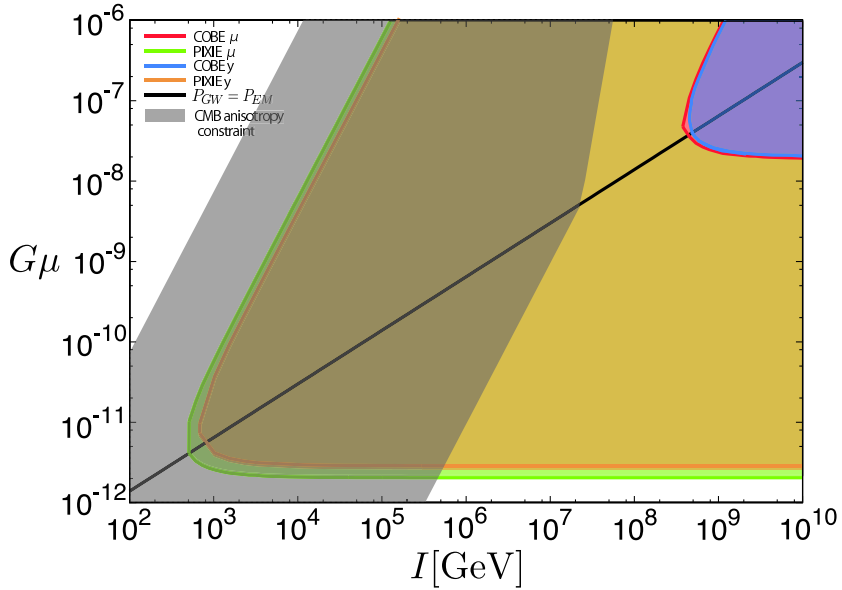}
\label{fig:CMBdistconstraintsmall}
}

\end{center}
\end{minipage}
\caption{The red and blue regions are excluded by $\mu$ and $y$ distortion measurement by COBE, respectively.
The green and orange regions will be probed by $\mu$ and $y$ distortion measurement by PIXIE, respectively.
The gray region is excluded by the CMB anisotropy constraint.}
\label{fig:CMBdistconstraint}
\end{figure}

SCS loops inject energy into the baryon-photon fluid in two ways; EMW emission and plasma dissipation.
The energy injected per unit time at time $t$ is given by
\be
\frac{dQ}{dt}(t)=(P_{\rm EM}+P_{\rm dis})n_d(t),
\ee
where $P_{\rm dis}\simeq F_{\rm dis} \times \alpha t_i(t)$ since we are considering only disappearing loops.
The chemical potential of the CMB photons is calculated by\cite{Hu:1992dc}
\be
\mu \approx 1.4\int^{t_{\mu,{\rm f}}}_{t_{\mu,{\rm i}}} dt \frac{dQ/dt}{\rho_\gamma}, \label{muSCS}
\ee
where $\rho_\gamma$ is the photon energy density and $t_{\mu,{\rm i}}$ and $t_{\mu,{\rm f}}$ are the times which correspond to the redshift $z=2\times 10^6$ and $z=5\times 10^4$, respectively.
The $y$ parameter generated by SCS loops is given by\cite{Hu:1992dc}
\be
y \approx \frac{1}{4}\int^{t_{y,{\rm f}}}_{t_{y,{\rm i}}} dt \frac{dQ/dt}{\rho_\gamma}, \label{ySCS}
\ee
where $t_{y,{\rm i}}=t_{\mu,{\rm f}}$ and $t_{y,{\rm f}}$ is the time at the recombination, $z\simeq 1090$.

The parameter sets $(G\mu,I)$ which are excluded by COBE FIRAS measurement or will be probed by PIXIE are shown in Fig.~\ref{fig:CMBdistconstraint}.
We also show the CMB anisotropy constraint.
The shapes of the excluded region and the region probed by PIXIE are similar to Fig.~\ref{fig:CMBconstraint}, since both types of constraint focus on the energy injection at some epoch.
However, the regions shown in Fig. \ref{fig:CMBdistconstraint} do not have the right edge as Fig.~\ref{fig:CMBconstraint}, which corresponds to the threshold value of the energy of injected photons.
Therefore, there are parameters excluded by not the CMB anisotropy constraint but the CMB spectral distortion constraint in the large $I$ region, although the excluded region by the distortion constraint where $I$ is smaller is mostly covered by the anisotropy constraint.

In the large-loop case of Fig.~\ref{fig:CMBconstraint}, under the dashed and dotted lines $({\tau}_{\rm dis}H)^{-1}>1$ 
over the epochs when $\mu$ and $y$ distortions are generated, respectively.
Under these lines, enhancement of loop number density does not occur, hence the observable CMB spectral distortion cannot be generated unless the energy density of infinite strings exceeds some threshold value.
This corresponds to $G\mu\gtrsim3\times 10^{-12}$.
In the small-loop case and when EMW emission or plasma dissipation dominates GW emission, the loop energy density, which is comparable with that of infinite string, is immediately converted into the injected energy.
Therefore, the lower bound of the parameter regions shown in Fig.~\ref{fig:CMBdistconstraintsmall} also correspond to the threshold value of the energy density of infinite strings.

\subsection{BBN constraint}

Photons emitted by SCS loops at BBN epoch might destroy produced light elements and affect their final abundances~\cite{Kawasaki:2004yh,Jedamzik:2004er}.
Since the observed abundances of light elements are basically consistent with the standard prediction of BBN, 
nonstandard energy injection from SCS loops is constrained.
In particular, the constraint from the abundance ratio of $^3{\rm He}$ to ${\rm D}$, 
which may be altered by the process $^4{\rm He}+\gamma \rightarrow {\rm n} + {^3}{\rm He}$, 
is most stringent in the photo-dissociation processes~\cite{Kawasaki:2004yh}.
Figure 42 in Ref.~\cite{Kawasaki:2004yh} shows the upper bound of the abundance of the decaying particle which causes photo-dissociation as a function of the lifetime of the particle, which can be easily translated into the bound
on the energy density of SCS loops.
The constraint on energy injection around $t\sim 10^8{\rm s}$ is especially stringent.
It should be noticed that, in order to destroy nuclei, photons must have energy larger than the threshold energy of the destruction process. Actually a photo-dissociation process 
$^4{\rm He}+\gamma \rightarrow {\rm n} +{^3}{\rm He}$ requires the photon energy larger than $20.6$\,MeV.
Therefore, we consider parameter sets where $\omega_c(t)=\mu^{3/2}I^{-3}(l_d(t))^{-1}>20.6\,{\rm MeV}$ is satisfied.
Otherwise, BBN constraints do not apply.

\begin{figure}
\begin{minipage}{1\hsize}
\begin{center}
\subfigure[The large-loop case, $\alpha=0.1$. Under the dashed line, $\dot{\tau}_{\rm dis}H^{-1}>1$ at $t=10^8{\rm s}$.]{
\includegraphics[width=110mm]
{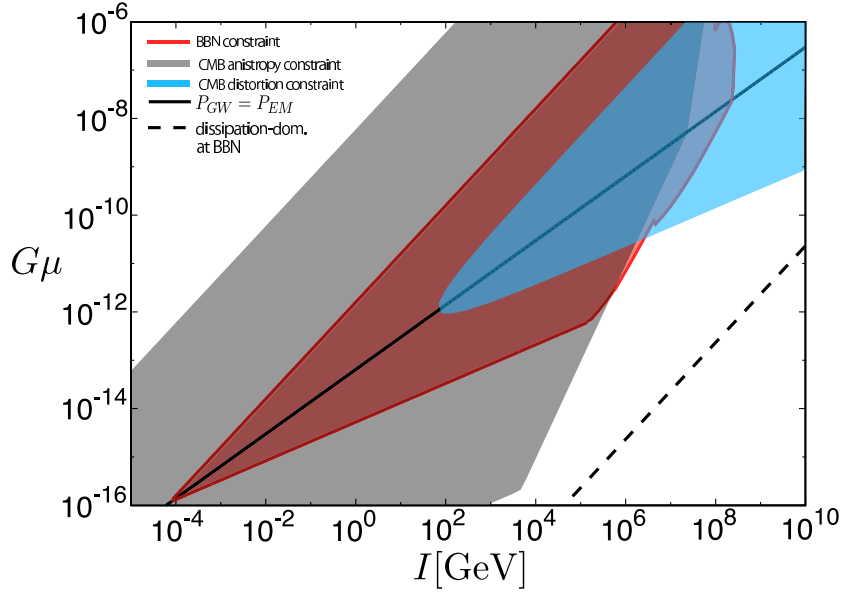}
\label{fig:BBNconstraintlarge}
}
\subfigure[The small-loop case, $\alpha=\Gamma_{\rm eff}$]{
\includegraphics[width=110mm]
{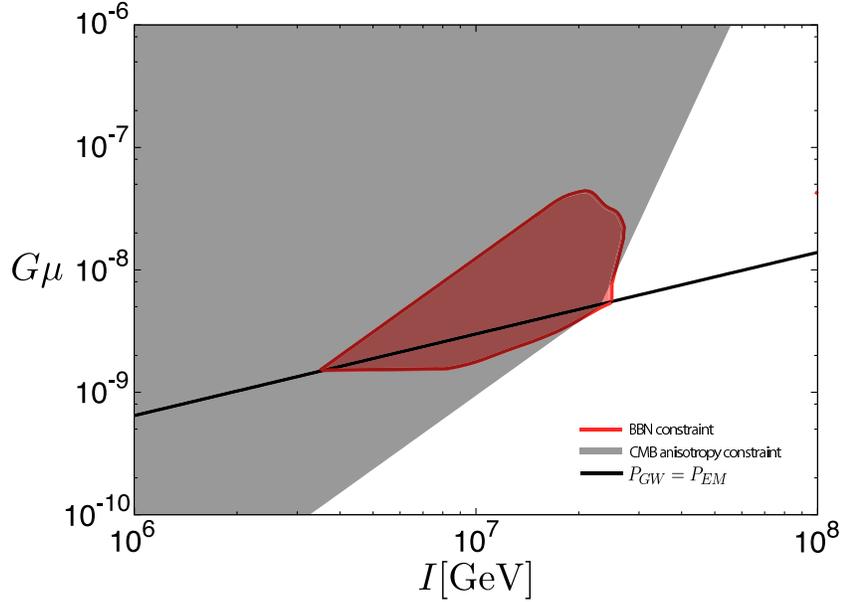}
\label{fig:BBNconstraintsmall}
}

\end{center}
\end{minipage}
\caption{The BBN constraint in the large-loop case (top panel) and the small-loop case (bottom panel).
In each figure, the red region represents the BBN constraint, the grey region is excluded by the CMB anisotropy constraint, the blue region is excluded by the CMB distortion constraint and on the black line the energy radiated by EMW emission per unit time is equal to that by GW emission.}
\label{fig:BBNconstraint}
\end{figure}

The resultant constraint is shown in Fig.~\ref{fig:BBNconstraint}.
Most of the excluded region comes from the energy injection around $t\sim 10^8\,{\rm s}$, 
when the energy injection is constrained most severely.
The right side of the region is cut because the injected photon energy becomes small for large $I$ and eventually
it falls below $20.6$\,MeV.

In terms of the ratio of the injected energy to the background energy, the BBN constraint is not so severe as the CMB anisotropy constraint.
For the BBN constraint, the upper bound is $\rho_{\rm inj}(t)/\rho_{\rm tot}(t) \lesssim 10^{-7}$ at $t\sim 10^8\,{\rm s}$, 
while the CMB anisotropy constraint requires $\rho_{\rm inj}(t)/\rho_{\rm tot}(t) \lesssim 10^{-11}$ at $t\sim 10^{13}\,{\rm s}$.
Therefore, the excluded region by the BBN constraint is smaller than that by the CMB anisotropy constraint.
However, in the large-loop case, there is a region which is excluded by not the CMB anisotropy constraint but the BBN constraint, since there is a region where SCS loops cannot emit photons with energy higher than $13.6\,{\rm eV}$ around recombination, but can emit photons with $\omega>20.6\,{\rm MeV}$ at BBN epoch.
Nevertheless, when combined with the CMB spectral distortion constraint, 
most of the excluded region by the BBN constraint is covered.
In the small-loop case, the region excluded by the BBN constraint is so small that it is covered by the CMB anisotropy constraint almost completely.
In Fig.~\ref{fig:BBNconstraintlarge}, we plot the dashed line under which $({\tau}_{\rm dis}H)^{-1}>1$ at $t=10^8\,{\rm s}$, but the whole excluded region in the figure is above this line.
In the small-loop case, such a line is below the range of Fig.~\ref{fig:BBNconstraintsmall}.
Therefore, plasma dissipation has nothing to do with the BBN constraint.

\subsection{Radio wave observations}

In the above two subsections, we have considered the constraint for SCSs decaying before/during recombination,
where emitted photons are mainly absorbed.
On the other hand, photons from SCS loops can reach us and be observed directly if they are emitted after the recombination.
First, let us consider the possibility that they are observed as radio wave.
While possible radio wave signatures from SCS loops are studied in Refs.~\cite{Vachaspati:2008su,Cai:2011bi,Cai:2012zd},
we here aim to derive the constraint on parameters from radio wave observations.

Considering only disappearing loops, the detection rate of radio wave bursts of frequency $f$ from SCS loops is calculated as
\be
\dot{N}(f)=\int^{z_{\rm rec}}_0 dz \frac{d\dot{N}}{dz}(z,f),
\label{Ndot}
\ee
where $d\dot{N}/dz$, the rate of bursts coming from redshift $z$ is given by
\begin{align}
\frac{d\dot{N}}{dz}(z,f)\sim n_d(z)\frac{dV}{dz}(z)\frac{1}{(1+z)^3} & \frac{c}{\alpha t_i(z)(1+z)} \frac{\theta_m(z,f)^2}{4}\frac{f_s}{4\pi} \nonumber \\
& \times \Theta(1-\theta_m(z,f))\Theta(S(z,f)-S_0)\Theta(f_c(z)-(1+z)f).
\label{dNdzradio}
\end{align}
Here, $z_{\rm rec}=1090$ is the redshift at recombination, $\frac{dV}{dz}(z)=4\pi r(z)^2\frac{dr}{dz}$
is the volume element, $r(z)=\int^z_0\frac{dz^{\prime}}{H(z)}$ is the comoving distance, 
$H(z)=H_0\sqrt{\Omega_{\Lambda}+\Omega_m(1+z)^3+\Omega_r(1+z)^4}$ is the Hubble parameter, $c$ is the number of the appearance of cusps on a loop per oscillation period, which is set to be $1$ in this paper, 
$\theta_m(z,f)=(\alpha t_i(z)f(1+z))^{-1/3}$ is the angle of the emitted photon from a cusp, 
$f_s$ is the sky coverage of the telescope, $S(z,f)$ is the energy flux per unit frequency interval
\footnote{
Hereafter, we call it simply as ``flux".
}
of a burst with frequency $f$ coming from redshift $z$, $S_0$ is the sensitivity of the telescope, 
$f_c(t)=\omega_c(t)/2\pi$ is the cutoff frequency and $\Theta(x)$ is Heviside step function.
The first step function in (\ref{dNdzradio}) is necessary since a cusp on a loop of length $l$ emit only EMWs of frequency $f\gg l^{-1}$, and the third one reflects the fact that the frequency of EMWs from SCS loops has the upper bound.
$S(z,f)$ is given by
\be
S(z,f)\sim\frac{dE}{df^{\prime}d\Omega}\frac{df^{\prime}}{df}\frac{1}{r(z,f)^2}\frac{1}{\Delta(z,f)}\frac{1}{1+z}\sim \frac{I^2(\alpha t_i(z))^2}{r(z)^2\Delta(z,f)},
\label{flux}
\ee
where $f^{\prime}=(1+z)f$ is the frequency of a burst when it was emitted.
The duration of a burst of frequency $f$ emitted at redshift $z$, $\Delta(z,f)$, is given by
\be
\Delta(z,f)=\max\left\{\sqrt{\Delta t_{\rm int}^2 + \Delta t_s^2}, \Delta_{\rm res}\right\}.
\ee
Here $\Delta t_{\rm int}$ is the duration of the burst event at the cusp times $1+z$ and given by~\cite{Babul:1986wd}
\be
\Delta t_{\rm int}  \sim (1+z)\Delta t_{f'}\theta_m(f')^2\sim  f^{-1},  \label{t_int}
\ee
where $\Delta t_{f'} \sim l\theta_m(f')$ is the emission time for a photon of frequency $f$ at a cusp,
and we have used the fact that the Lorentz factor of a cusp, which is directed toward us, is roughly given by $\theta_m(f')^{-1}$.
The time delay due to scattering by the turbulent inter-galactic medium is expressed by $\Delta t_s$,
which is given by~\cite{Lorimer:2007qn,Kulkarnietal,Cai:2011bi}
\be
\Delta t_s=\Delta t_1\left(\frac{1+z}{1+z_1}\right)^{1-\beta}\left(\frac{f}{f_1}\right)^{-\beta},
\ee
where we set $\Delta t_1=5{\rm ms}, z_1=0.3,f_1=1.374\,{\rm GHz}, \beta=4.8$~\cite{Cai:2011bi}.
Finally, $\Delta_{\rm res}$ is the time resolution of the telescope.
Since $\Delta t_{\rm int}\ll \Delta t_s$, $\Delta=\Delta t_s$ if $\Delta t_s>\Delta_{\rm res}$ and $\Delta=\Delta_{\rm res} $ if $\Delta t_s<\Delta_{\rm res}$.

According to Ref.~\cite{Lorimer:2007qn}, only one burst-like event was discovered in the analysis of 20 days data 
which was collected at Parkes observatory.
Assuming that the number of bursts detected in a period obeys Poisson distribution, we can calculate the upper bound of the detection rate of bursts at Parkes observatory as $\dot N_{\lim}=96\,{\rm yr}^{-1}$ at 95\% C.L..
The sensitivity, time resolution and sky coverage of Parkes telescope when the data was recorded, are $S_0=0.3\,{\rm Jy}, \Delta_{\rm res}=1\,{\rm ms}, f_s=3.1\,\times 10^{-3}{\rm sr}$~\cite{Lorimer:2007qn}, respectively.
Therefore, setting the parameter about the telescope as above and the frequency as $f=1.4\,{\rm GHz}$, to which the Parkes telescope is sensitive, we search for parameter sets for which $\dot N(f=1.4\,{\rm GHz})$ exceeds $\dot N_{\lim}$.
We also consider prospects for SCS searches in on-going radio wave observations.
As an example, we take the High Time Resolution Universe (HTRU) survey at Parkes~\cite{Keith:2010kk}.
The parameters about the survey are set as $S_0=0.61\,{\rm mJy}, \Delta_{\rm res}=64\,\mu {\rm s}, f_s=3.1\times 10^{-3}\,{\rm sr}, f=1.4\,{\rm GHz}$~\cite{Keith:2010kk,BurkeSpolaor:2011rp}.

\begin{figure}[tp]
\begin{minipage}{1\hsize}
\begin{center}
\subfigure[The large-loop case, $\alpha=0.1$. Under the black dotted line, $({\tau}_{\rm dis}H)^{-1}>1$ at the recombination.]{
\includegraphics[width=95mm]
{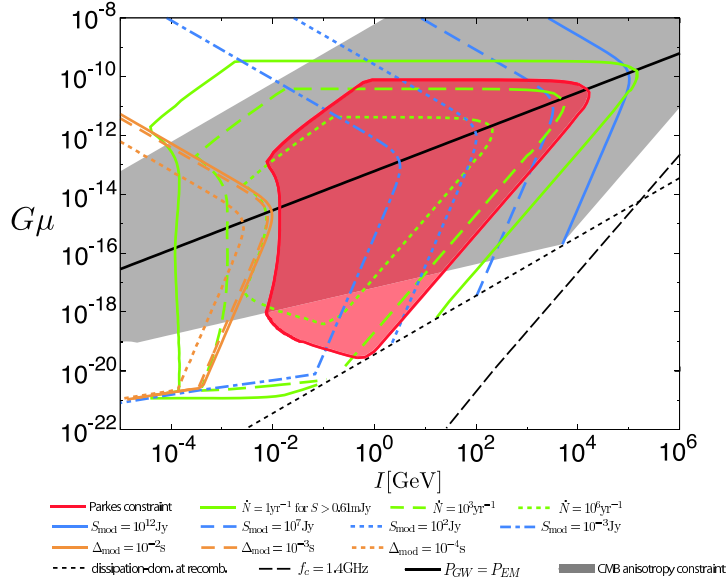}
\label{fig:radioconstraintlarge}
}
\subfigure[The small-loop case, $\alpha=\Gamma_{\rm eff}$.]{
\includegraphics[width=95mm]
{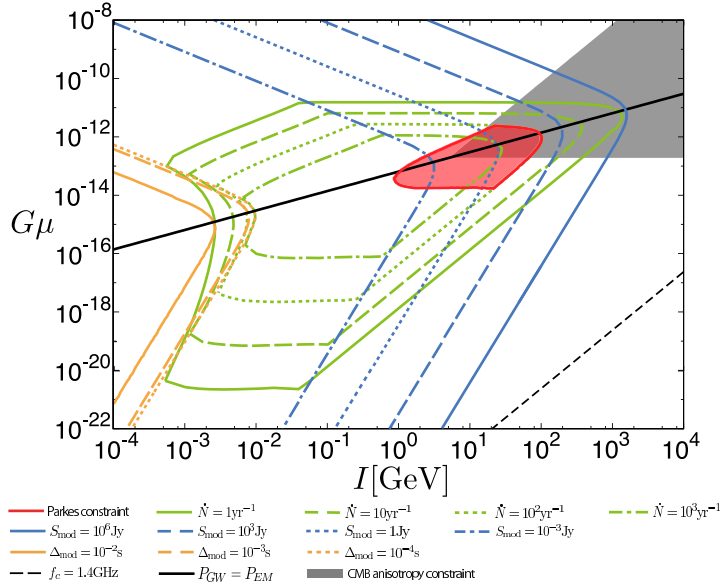}
\label{fig:radioconstraintsmall}
}

\end{center}
\end{minipage}
\caption{Constraint from current radio wave observations and the prospect for future ones 
in the large-loop case (top panel) and the small-loop case (bottom panel).
In each panel, the red region represents the excluded region by the past data of the observation at Parkes.
Green lines are the contours of the rate of burst with frequency $f=1.4\,{\rm GHz}$ and flux larger than $0.61\,{\rm mJy}$, which can be detected in Parkes HTRU.
Blue (Orange) lines are contour plots of the mode flux (duration) of such bursts.
Below the black dashed line, radio waves with frequency $f=1.4\,{\rm GHz}$ cannot be emitted by SCS loops after recombination.
The shaded region is excluded by the CMB anisotropy constraint.
}
\label{fig:radioconstraint}
\end{figure}

 The result is shown in Fig.~\ref{fig:radioconstraint}.
 In this figure, the constraint from past data of Parkes, contours for the detection rate, mode value of flux $(S_{\rm mod})$ 
 and duration of bursts $(\Delta_{\rm mod})$ in Parkes HTRU survey are shown, in addition to the CMB anisotropy constraint.
 Here $S_{\rm mod}$ and $\Delta_{\rm mod}$ are defined as $S(z_m,f)$ and $\Delta(z_m,f)$, 
 where $z_m$ is the redshift for which $d\dot N/d\ln z(z,f)$ becomes maximum.
  This figure shows that although the constraint from the past Parkes data is considerably covered by the CMB anisotropy constraint, 
 there are parameter sets excluded by the Parkes data but consistent with CMB anisotropy constraint, 
 especially in the case where EMW emission dominates GW emission.
 
The Parkes HTRU survey largely extends the parameter region which can be explored.
We here comment on the contours of the detection rate, $S_{\rm mod}$ and $\Delta_{\rm mod}$ in Fig.~\ref{fig:radioconstraint}.
The flux $S(z,f)$ is the decreasing function of $z$, while $dN/d\ln z(z,f)$ is the increasing function, as long as $S(z,f)>S_0$.
Therefore, the largest contribution to the integral (\ref{Ndot}) comes from largest $z$.
In other words, most of detectable bursts come from the maximum redshift at which bursts large enough to be detected can be emitted.
In Fig.~\ref{fig:radioconstraint}, on the right side of the line of $S_{\rm mod}=10^{-3}\,{\rm Jy}$, 
$S(z_{\rm rec},f)$ exceeds $S_0=0.61\,{\rm mJy}$ and hence most bursts come from $z\sim z_{\rm rec}$.
Note that for larger $I$ the flux of burst becomes larger, as easily found from (\ref{flux}).
The burst rate on the right of the $S_{\rm mod}=10^{-3}{\rm Jy}$ line is basically determined by the number density and the oscillation period of disappearing loops at around the recombination epoch, $n_d(z_{\rm rec})$ and $\alpha t_i(z_{\rm rec})$. 
On the other hand, on the left side of the line of $S_{\rm mod}=10^{-3}{\rm Jy}$, $S(z_{\rm rec},f)$ is smaller than $S_0$ 
and most detectable bursts come from the redshift $z_m (< z_{\rm rec})$ which satisfies $S(z_m,f)=S_0$.
In this region, most bursts have flux comparable with $S_0$ and the burst rate is determined by $n_d(z_m)$ and $\alpha t_i(z_m)$.
When $z_m\lesssim 1$, $\Delta t_s(z_m,f)$ is larger than $\Delta_{\rm res}$, then $\Delta_{\rm mod}=\Delta t_s(z_m,f)$.

We here also mention to the effect of plasma dissipation in the large-loop case.
Under the black dotted line in Fig. \ref{fig:radioconstraintlarge}, there is a dissipation-dominated epoch which ends at the recombination.
Therefore, for parameter sets under this line, the radio burst rate is strongly suppressed since all loops which can emit radio wave bursts are formed only after the recombination and the loop number density is not enhanced.
Even above this line, observation of radio wave burst is affected by plasma dissipation for $G\mu \lesssim 10^{-20}$, since there is a dissipation-dominated era 
and loops which emit radio wave bursts are formed at the end of that epoch.
In the small-loop case, loops which exist at time $t$ were generated around $t$.
Although there are parameter sets for which radio bursts emitted after the reionization are dominant among all bursts, dissipation is not important for such parameter sets since $I$ is so small.
Therefore, we can neglect plasma dissipation in the small-loop case.

We give a comment on the radio wave background.
Although the radio wave background has been observed by various telescopes, radio waves from SCS loops cannot be a part of this.
This is because SCS loops emit radio wave bursts, whose duration is much less than the time interval of arrivals of bursts.
Since the duration of radio bursts are shorter than the time interval of each event, i.e. $d\dot N/d\ln z(z,f)\times \Delta(z,f) <1$,\footnote{
A similar criterion was adopted in the calculation of the GW background which consists of GW bursts from cosmic string loops~\cite{Damour:2000wa}.
}
radio bursts from SCS loops should not be included in the diffuse background.
Therefore, there is no constraint on SCSs from observations of the radio background.

\subsection{X-ray and $\gamma$-ray observations}


We here focus on the possibility that high energy photons, such as X-rays or $\gamma$-rays, emitted by SCS loops are detected.
Such photons can be detected telescopes such as Fermi~\cite{Fermi}, on which two instruments are loaded.
One is Gamma-Ray Burst Monitor (GBM)~\cite{Meegan:2009qu}
and the other is Large Area Telescope (LAT)~\cite{Atwood:2009ez}.
Let us consider whether they can detect high-energy photons from SCS loops or not.

First, we consider how the high-energy photon events caused by SCS loops are observed with Fermi.
Since there is no time-delay effect for high-energy photons unlike radio waves, duration of a burst of photons with energy 
$\omega\gtrsim 1\,{\rm keV}$ is simply given by
$\Delta \sim \omega^{-1} \sim 10^{-18}\,{\rm s}\times\left(\omega/{\rm keV}\right)^{-1}$ (see Eq.~(\ref{t_int})).
This is much shorter than the time resolution of the detector on Fermi($\sim \mu {\rm s}$ for GBM~\cite{Meegan:2009qu} and $\sim 10\,\mu {\rm s}$ for LAT~\cite{Atwood:2009ez}).
Therefore, when photons from a SCS loop come to Fermi, they all enter the detector in a time shorter than the time resolution, 
hence there are no incident photons in the next time bin.
On the other hand, it is almost impossible that multiple photons of the diffuse background enter the detector in the time resolution 
and the astrophysical events such as GRBs last much longer than the time resolution.
We therefore regard such instant detections as unique events of SCSs, and take a criterion that 
a SCS event is detected if many photons enter the effective area of the detector in one event.

Then we can calculate the detection rate of high-energy photon events of SCSs as
\be
\dot{N}=\int^{z_{\rm rec}}_0dz\int \frac{d^2\hat{n}_b}{4\pi} \int \frac{d^2\hat{n}_F}{4\pi}\frac{d\dot{N}_{b}}{dz}(z)P(z,\hat{n}_b,\hat{n}_F),
\label{dotNgammaX1}
\ee
\be
\frac{d\dot{N}_{b}}{dz}(z)= n_d(z)\frac{dV}{dz}(z)\frac{1}{(1+z)^3}\frac{c}{\alpha t_i(z)(1+z)},
\ee
\be
P(z,\hat{n}_b,\hat{n}_F)=\Theta(N_{\gamma}(z,\theta_b) - 2)\Theta\left(\arccos\left(1-\frac{f_{s}}{2\pi}\right)-\theta_F\right), \label{Pdet}
\ee
\be
N_{\gamma}(z,\theta_b)= \int^{\max\{\omega_l,\min\{ \omega_{\rm th}(z,\theta_b),\omega_{h}\}\}}_{\omega_{l}}d\omega 
\frac{dN_\gamma}{d\omega}(z,\theta_b),
\label{Ngamma}
\ee
\begin{align}
\frac{dN_\gamma}{d\omega}(z,\theta_b)= \frac{dn}{d\omega^\prime}(z,\omega^{\prime})\frac{d\omega^{\prime}}{d\omega}& \frac{A}{\pi(\theta_m(z,\omega/2\pi))^2(a_0r(z))^2}\nonumber \\
&\times \Theta(1-\theta_m(z,\omega/2\pi))\Theta(\omega_c(z)-\omega^\prime)\Theta_a(z,\omega),
\end{align}
The meaning of each symbol is as follows.
$d\dot{N}_{b}/dz$ is the occurrence rate of EMW bursts at redshift $z$.
$P(z,\hat{n}_b,\hat{n}_F)$ is the probability that a burst at redshift $z$ is detected,
where the unit vectors $\hat{n}_b$ and $\hat{n}_F$ represent the directions of the burst and the direction of the detector on Fermi faces, respectively.
The first step function in Eq.~(\ref{Pdet}) reflects the aforementioned detection criterion and the second one represents the condition that a SCS loop is in the field of view of detector.
Here, $\theta_b$($\theta_F$) is the angle between $\hat{n}_b$($\hat{n}_F$) and the line which connects the loop and Fermi,
$f_s$ is the sky coverage of the instrument,
and $N_{\gamma}(z,\theta_b)$ is the number of photons which come to Fermi in an event.
In Eq. (\ref{Ngamma}), $\omega_{\rm th}(z,\theta)=2\pi(1+z)^{-1}(\alpha t_i(t))^{-1}\theta^{-3}$
is the photon frequency at the emission with emission angle $\theta$,
$\omega_h$ and $\omega_l$ are the highest and lowest energy of photons which can be detected by the instrument,
$\omega^{\prime}=(1+z)\omega$ is the energy of the photon at its emission,
$A$ is the effective area of the detector,
and $dn/d\omega$ is the number of photons with energy $\omega$ from a cusp event, which is given by (see Eq.~(\ref{dE_df}))
\be
\frac{dn}{d\omega}(z,\omega)\sim \frac{1}{\omega}\frac{dE}{d\omega}\sim
I^2(\alpha t_i(z))^{4/3}\omega^{-5/3}.
\ee
The step functions in (\ref{Ngamma}) are inserted for the same reasons as (\ref{dNdzradio}).
$\Theta_a$ is introduced in order to take into account the optical depth of high-energy photons.
High-energy photons emitted at high redshift may scatter with background particles and not be able to reach the earth.
The transparency window for photons with energy $\omega$ emitted at redshift $z$ is summarized e.g. in Ref.~\cite{Chen:2003gz}.
We simply take $\Theta_a(z,\omega)=0$ if $(z,\omega)$ is in the black region of Fig.~2 of Ref.~\cite{Chen:2003gz}, 
otherwise $\Theta_a(z,\omega)=1$.
Eq.~(\ref{dotNgammaX1}) is simplified as
\be
\dot{N}=\int^{z_{\rm rec}}_0dz \frac{d\dot N}{dz}(z)=\int^{z_{\rm rec}}_0dz \frac{d\dot{N}_{b}}{dz}\frac{1}{2}[1-\cos(\tilde{\theta}_b(z))]\frac{f_s}{4\pi},
\label{dotNgammaX2}
\ee
where $\tilde{\theta}_b(z)$ is the solution of $N_{\gamma}(z,\tilde{\theta}_b)=2$ if it exists, otherwise 0.
$\tilde{\theta}_b(z)$ is the largest value of $\theta_b$ for which Fermi can detect multiple photons in the event.
In order for $N_\gamma$ to be nonzero, at least $\omega_{\rm th}(z,\theta_b)$ must exceed $\omega_l$.
This leads to $\tilde{\theta}_b\ll 1$, since $\omega_l\gtrsim {\rm keV}$ while $\alpha t_i(z)$ is cosmological. 
Hereafter we take parameters about the instrument  as below.
\be
A=100\,{\rm cm}^2, \omega_{l}=20\,{\rm keV}, \omega_{h}=10\,{\rm MeV},f_{s}=4\pi 
\ee
for GBM\cite{Meegan:2009qu}, and
\be
A=9500\,{\rm cm}^2, \omega_{l}=20\,{\rm MeV}, \omega_{h}=300\,{\rm GeV},f_{s}=2.4
\ee
for LAT\cite{Atwood:2009ez}.

\begin{figure}[tp]
\begin{center}
\includegraphics[width=105mm]
{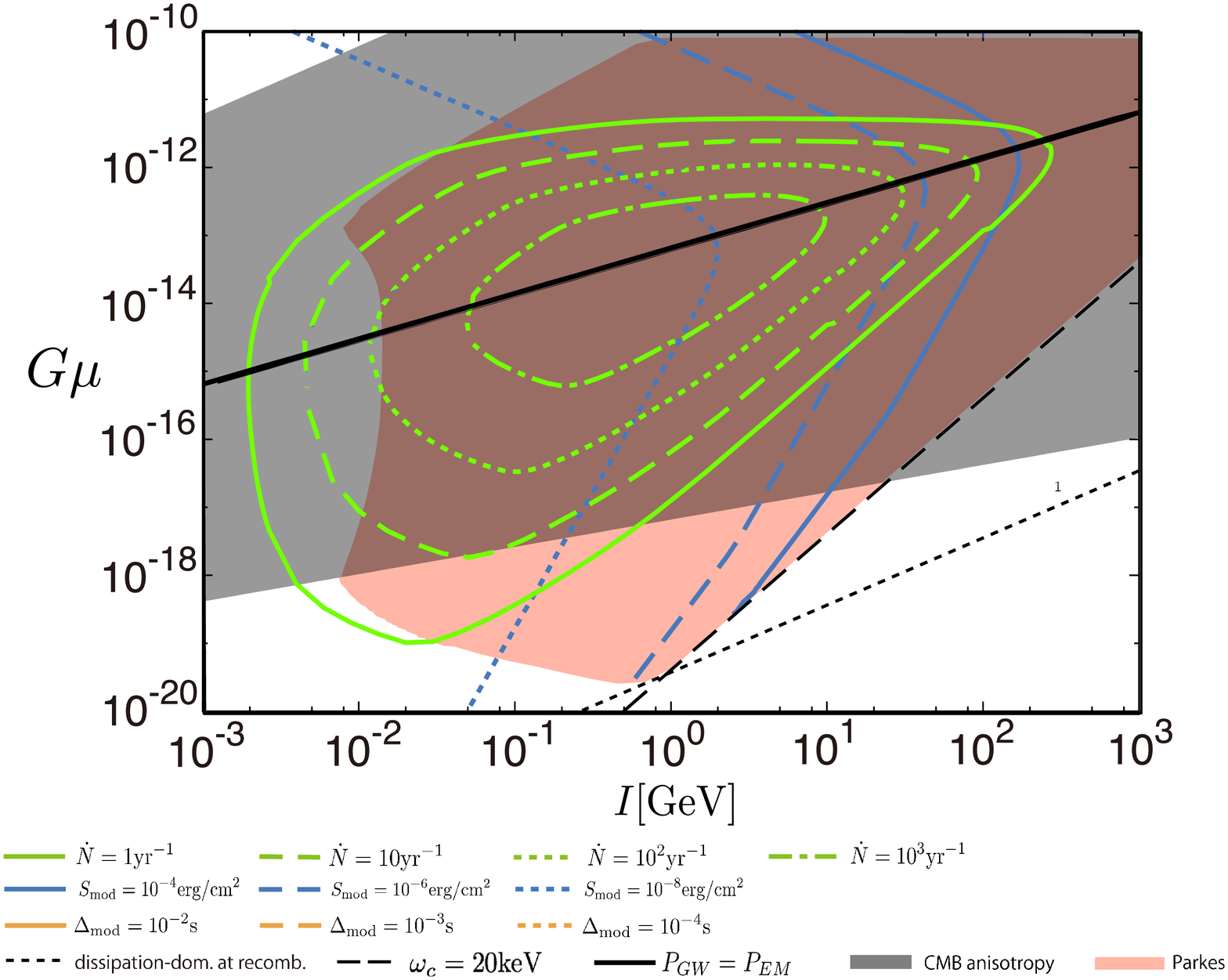}
\caption{
Contour plots of the detection rate (green lines) and the mode value of the fluence (blue lines) 
at the Fermi GBM. Here $\alpha$ is taken to be $0.1$.
Below the black dashed line, SCS loops cannot emit photons with energy $20\,{\rm keV}$ after recombination. 
The grey and red regions are excluded by the CMB anisotropy constraint and by the past data of Parkes, respectively.
Under the black dotted line, $({\tau}_{\rm dis}H)^{-1}>1$ at the recombination.}
\label{fig:Xconstraint}

 ~ \\

\includegraphics[width=105mm]
{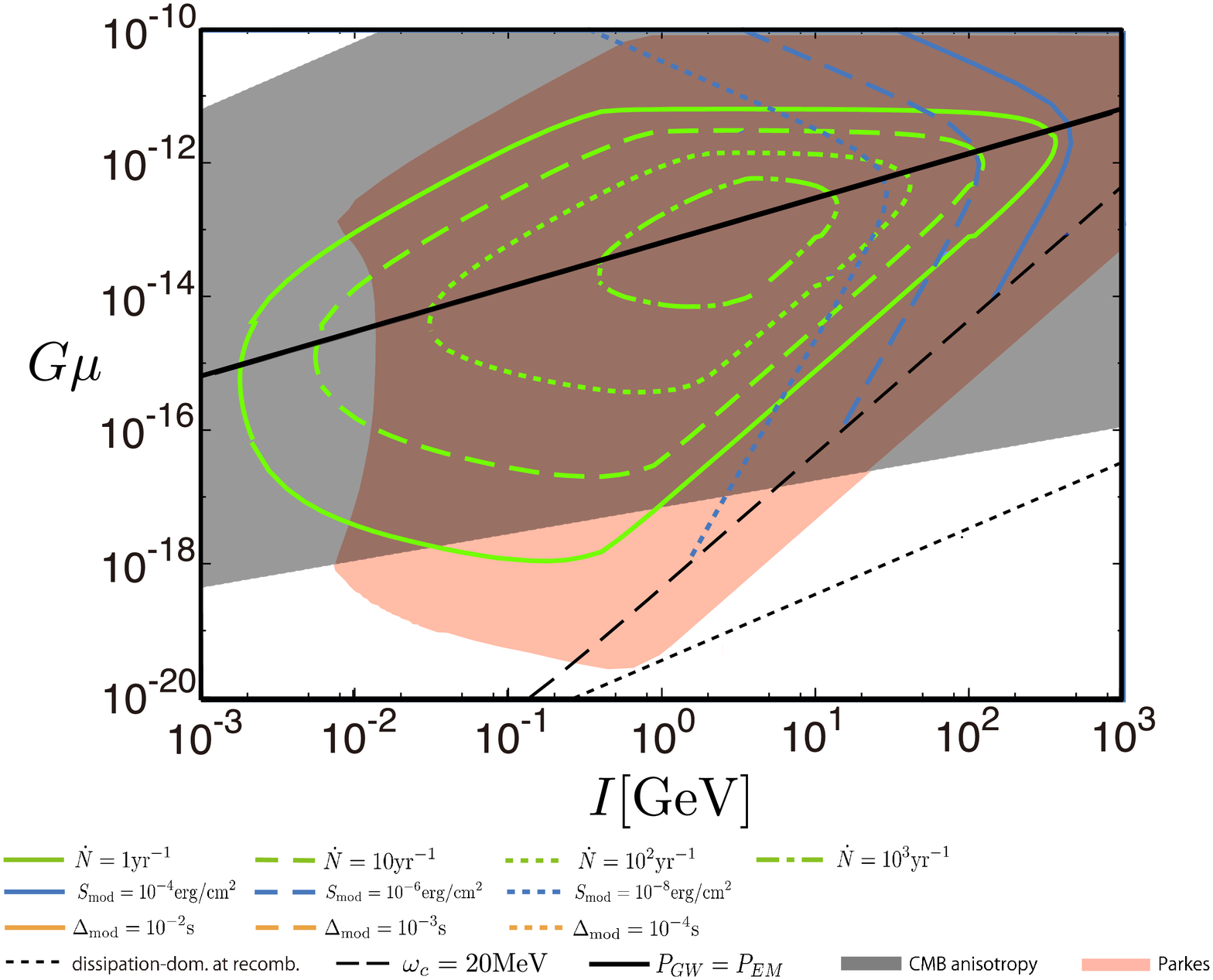}
\caption{
Contour plots of the detection rate (green lines) and the mode value of the fluence (blue lines) at the Fermi LAT.
Here $\alpha$ is taken to be $0.1$.
Below the black dashed line, SCS loops cannot emit photons with energy $20\,{\rm MeV}$ after recombination. 
The meanings of grey and red regions and the black solid and dotted lines are same as in Fig.~\ref{fig:Xconstraint}.
}
\label{fig:gammaconstraint}
\end{center}
\end{figure}

Then we derive the contours of the detection rate of unique events of SCSs in X-ray observation with GBM and $\gamma$-ray observation with LAT.
The results are shown in Fig.~\ref{fig:Xconstraint} for GBM and in Fig.~\ref{fig:gammaconstraint} for LAT.
Only the results in the large-loop case are shown, since in the small-loop case there is no parameter set for which events are detected at the rate higher than $1\,{\rm yr}^{-1}$ with both GBM and LAT.
We also plotted contours of the mode value of fluence, the total energy going through the unit area on the detector in an event.
The fluence of a burst at redshift $z$ with direction $\hat{n}_b$ is given by 
\be
S(z,\theta_b)= \frac{1}{A}\int^{\max\{\omega_l,\min\{ \omega_{th}(\theta_b),\omega_{h}\}\}}_{\omega_{l}}d\omega ~\omega 
\frac{dN_\gamma}{d\omega}(z,\theta_b),
\ee
We define $S_{\rm mod}$, the mode value of $S(z_m,\tilde\theta_b(z_m))$, where $z_m$ is the redshift at which $d\dot N/d\ln z(z)$ becomes maximum.
In the large-loop case, there is a parameter region where the event rate is more than $1\,{\rm yr}^{-1}$, but  all of the region is covered by the CMB anisotropy constraint or the constraint from the radio burst observation at Parkes for LAT.
For GBM, there is a small region which is not excluded by other constraints and where $\dot N\sim 1\,{\rm yr}^{-1}$.
Note that, similar to the flux of radio bursts, $N_\gamma(z,\theta_b)$ is a decreasing function of $z$ and an increasing function of $I$.
The region where $\dot N\sim 1\,{\rm yr}^{-1}$ with avoiding other constraints
in Fig.~\ref{fig:Xconstraint} corresponds to parameter sets for which the number of photons detected 
in an event from $z\sim z_{\rm rec}$ is much smaller than 1.
In this region, most detectable events come from $z\sim 1$ and the mode value of fluence is around $6\times 10^{-10}\,{\rm erg}/{\rm cm}^2$, which corresponds to the two photons with energy $\sim \omega_l =20\,{\rm keV}$.

In Figs.~\ref{fig:Xconstraint} and \ref{fig:gammaconstraint}, we also plotted the lines under 
which $({\tau}_{\rm dis}H)^{-1}>1$ at the recombination.
Since all parameter sets for which X-ray or $\gamma$-ray signals are detected at the considerable rate are above these line, plasma dissipation does not affect X-ray and $\gamma$-ray observations.
There is no parameter region for which the earlier dissipation-dominated epoch affects observations. 

Similar to the case of radio waves, X-rays and $\gamma$-rays emitted from SCS loops cannot contribute to the diffuse background because of the extremely short duration of EMW bursts.
Therefore, there is no constraint from the diffuse X-ray or $\gamma$-ray background.

\subsection{Pulsar timing limit on the stochastic GW background}

Like ordinary cosmic strings, SCSs emit GWs and a part of them forms stochastic GW background.
The amplitude of the GW background is usually written in the form of $\Omega_{\rm GW}(f)\equiv \frac{1}{\rho_{\rm cr}}\frac{d\rho_{\rm GW}(f)}{d\ln f}$, where $d\rho_{\rm GW}(f)/df$ is the energy density of GWs with frequency $f$ 
and $\rho_{\rm cr}$ is the critical density.
A formalism to calculate the GW spectrum $\Omega_{\rm GW}$ induced by cosmic string loops, where plasma dissipation is not taken into account, is given by~\cite{Damour:2000wa,Siemens:2006yp}
\be
   \Omega_{\rm GW}(f)=\frac{2\pi^2}{3H_0^2}f^3
   \int^{h_*}_0 dhh^2\int^{\infty}_0 dz \frac{d^2R}{dhdz}(f,h,z),
   \label{Omega_GW}
\ee
where $d^2R/dhdz$ is the arrival rate of bursts with amplitude $h$ and frequency $f$ emitted at redshift $z$ and $h_*$ is defined as
\be
\int^{\infty}_{h_*}dh\int^{z_*}_0 dz \frac{d^2R}{dhdz}(f,h,z) =f.\label{def_of_hstar}
\ee
GW bursts with amplitude $h<h_*$ overlap with each other, so we can think of them as a component of the stochastic background.
We can write $d^2R/dhdz$ as follows.
Similar to (\ref{dNdzradio}), the arrival rate of bursts emitted at redshift $z$ by loops with length $l$ is
\be
   \frac{d^2R}{dzdl}(f,z,l)
   = \frac{dn}{dl}(z,l)\frac{dV}{dz}(z)\frac{1}{(1+z)^3} \frac{2c}{l(1+z)}
   \frac{\theta_m^2(f,z,l)}{4}\Theta(1-\theta_m(f,z,l)).
   \label{dR_dzdl}
\ee
Here $dn/dl$ denotes the number density of loops with length $l$ and it is given by
\be
\frac{dn}{dl}(z,l)=\frac{1}{\alpha+\Gamma_{\rm eff}}\frac{dn}{dt_i}(t(z),t_i)\left(\frac{a(t_i)}{a(t(z))}\right)^3,
\ee
where $t_i$ is related to $l$ as Eq. (\ref{l_t}),
$\theta_m(f,z,l)=[(1+z)fl]^{-1/3}$,
and $c$ denotes the number of cusps which appear per oscillation period of a loop and is set to $1$ in this paper.
The amplitude of a GW burst with frequency $f$ emitted by a loop with length $l$ at redshift $z$ is given by
\be
   h(f,z,l)\simeq 2.68 \frac{G\mu l}{((1+z)fl)^{1/3}}\frac{1}{fr(z)}.
   \label{strain}
\ee

\begin{figure}[t]
\begin{tabular}{cc}
\begin{minipage}{0.5\hsize}
\begin{center}
  \subfigure[ the large-loop case, $\alpha=0.1$]{
\includegraphics[width=75mm]
{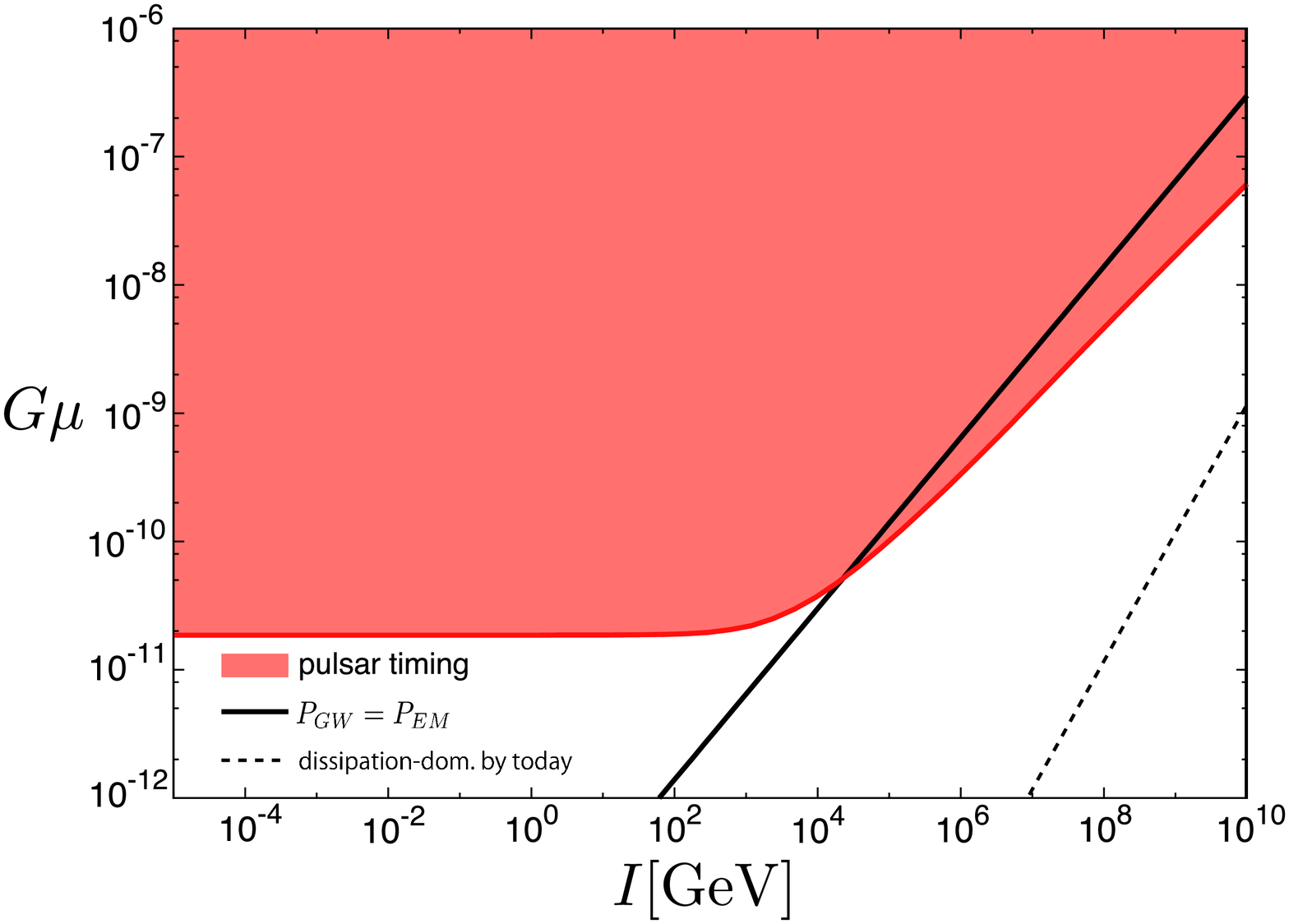}
\label{fig:pulsar_large}
}
\end{center}
\end{minipage}

\begin{minipage}{0.5\hsize}
\begin{center}
  \subfigure[ the small-loop case, $\alpha=\Gamma_{\rm eff}$]{
\includegraphics[width=75mm]
{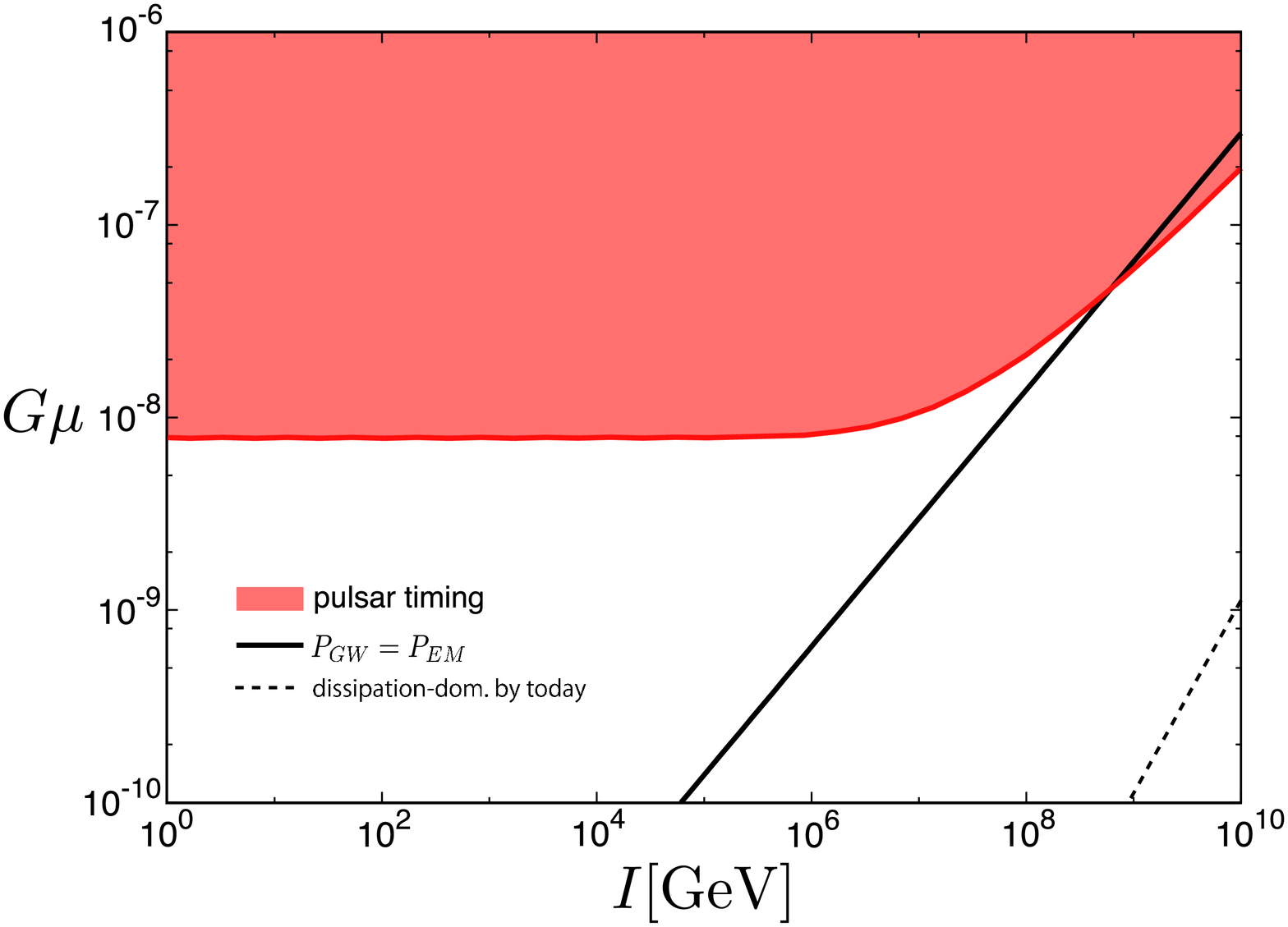}
\label{fig:pulsar_small}
}
\end{center}
\end{minipage}
\end{tabular}
\caption{Constraints from current pulsar timing experiments in the large-loop case (left panel) and the small-loop case (right panel).
Red regions are excluded.
On the black line the energy radiated by EMW emission per unit time is equal 
to that by GW emission.
}
\label{fig:pulsar}
\end{figure}

Pulsar timing experiments impose a severe constraint on the amplitude of the GW background at frequency $f\sim 1\,{\rm yr}^{-1}$.
One of the recent experiments NANOGrav~\cite{Demorest:2012bv} gives an upper limit as $\Omega_{\rm GW}<1.9\times 10^{-8}$ 
for $f\simeq 1/(5\,{\rm years})$.
We use this as the current constraint on the GW background induced by cosmic strings.

The resultant constraints in $I$-$G\mu$ plane are shown in Fig.~\ref{fig:pulsar} for the large-loop case and the small-loop case.
Again, the larger parameter region is excluded in the large-loop case than the small-loop case.
If GW emission dominates over EMW emission, a loop releases most energy as GWs and the energy density of such GWs is almost independent of $I$.
On the other hand, if EMW emission dominates over GW emission, $I$ determines 
the abundance of loops through the lifetime of loops and the ratio of energy emitted as GWs to that as EMWs.
Therefore, $I$ affects the spectrum of the GW background and the upper bound on $G\mu$ from the pulsar timing experiment becomes weaker for larger $I$ in this case.

In Fig.~\ref{fig:pulsar}, we also plotted the line above which there is no dissipation-dominated epoch in the history.
Since the excluded region in Fig.~\ref{fig:pulsar} is above this line, the above calculation in which plasma dissipation is neglected is justified.

\section{Summary and Discussion}

\begin{figure}[tp]
\begin{center}
\includegraphics[width=110mm]
{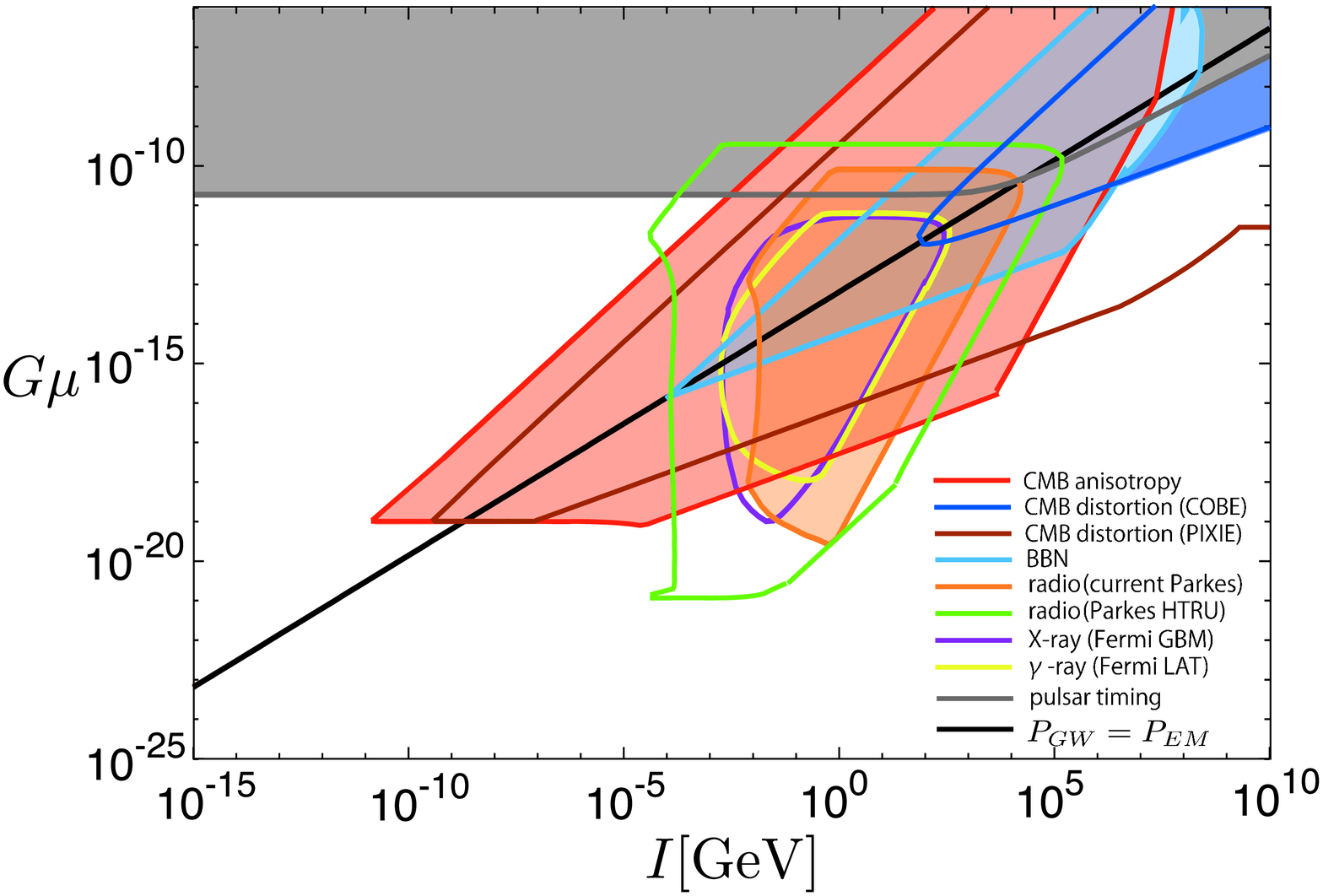}
\caption{
Summary of constraints in the large-loop case.
The red, light blue, blue and orange regions are excluded by CMB anisotropy measurement, CMB distortion measurement, BBN, pulsar timing experiments and the past observation of radio bursts at Parkes, respectively.
Inside the brown line, the CMB distortion will be detected by PIXIE.
Inside the green line, radio bursts will be detected at the rate higher than $1\,{\rm yr}^{-1}$ at the Parkes HTRU survey.
Inside the purple line, X-ray events are detected at the rate higher than $1\,{\rm yr}^{-1}$ at the Fermi GBM.
Inside the yellow line, $\gamma$-ray events are detected at the rate higher than $1\,{\rm yr}^{-1}$ at the Fermi LAT.
On the black solid line, the energy emitted as GWs from a SCS loops is equal to that as EMWs.
}
\label{fig:alllarge}

 ~ \\

\includegraphics[width=110mm]
{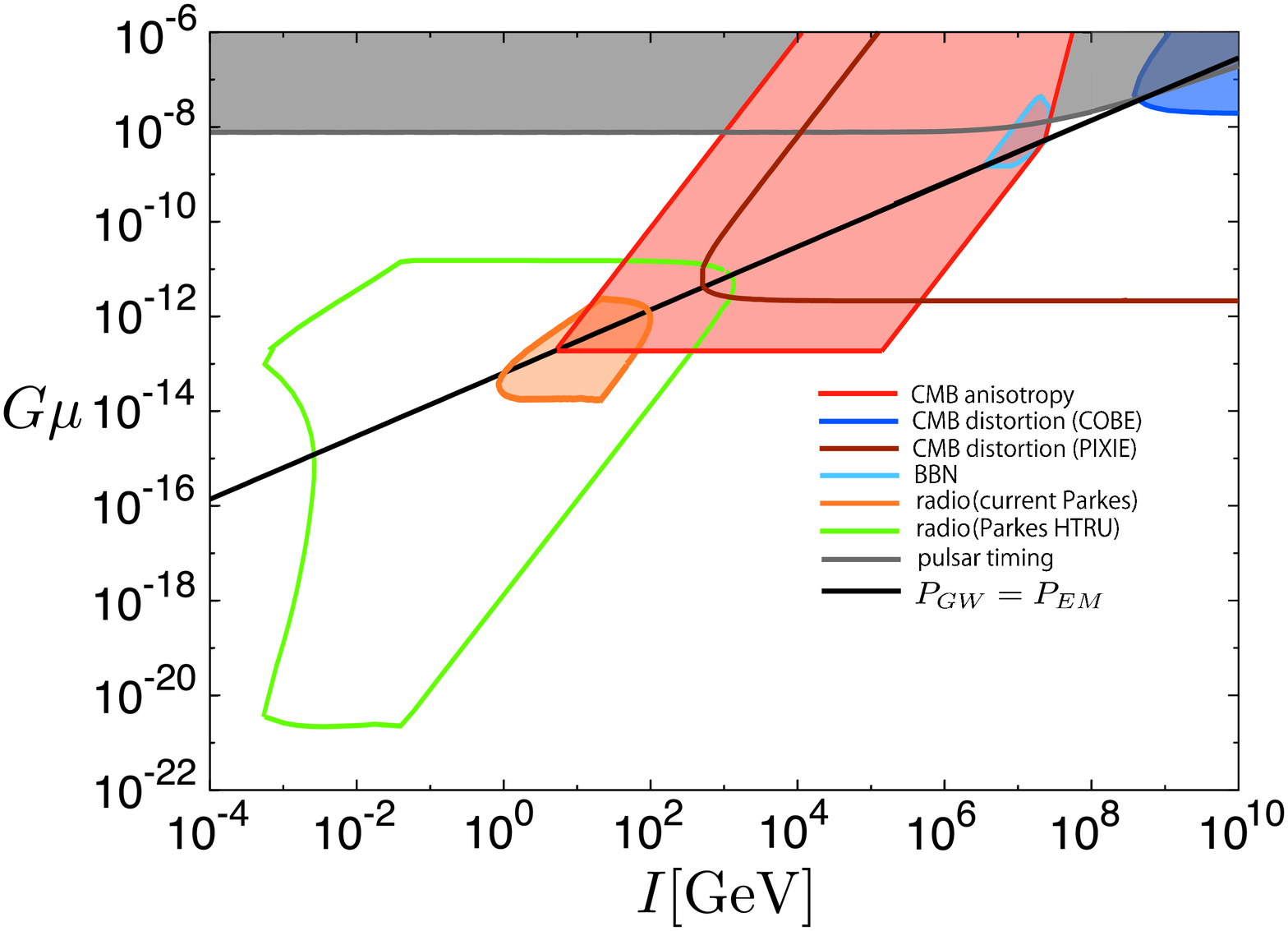}
\caption{
The same figure as Fig.~\ref{fig:alllarge}, but in the small-loop case.
There is no parameter regions for which X-ray ($\gamma$-ray) events are detected at the rate 
higher than $1\,{\rm yr}^{-1}$ with Fermi GBM (Fermi LAT).
}
\label{fig:allsmall}
\end{center}
\end{figure}

We explored various cosmological and astrophysical constraints on the parameters of SCS loops.
Results are summarized in Fig.~\ref{fig:alllarge} for the large-loop case and in Fig.~\ref{fig:allsmall} for the small-loop case.
These figures show that the CMB anisotropy constraint is most stringent and excludes the largest area in the parameter plane among various types of constraints.
However, there are some regions which are consistent with the CMB anisotropy observation but excluded by the CMB distortion constraint, the BBN constraint 
or the constraint from past Parkes data.
Measurements of the GW background in pulsar timing experiments also put upper bound on $G\mu$ for arbitrary $I$ in the case that GW emission dominates over EMW emission.
Therefore, we can say that different types of observations are complementary to each other for constraining SCSs.
Besides, we find that exploration of radio wave bursts in Parkes HTRU survey, one of the on-going high-sensitivity surveys can constrain SCSs or find the signature of them for large parameter regions.
It will be fruitful for the study of SCSs to perform the data analysis like Ref.~\cite{Lorimer:2007qn} in the on-going or future radio wave observations at not only Parkes HTRU survey but also LOFAR~\cite{LOFAR}, SKA~\cite{SKA} and so on.
Unlike radio wave observations, it is difficult to detect X-rays or $\gamma$-rays emitted by SCSs at high rate with Fermi.

\section*{Acknowledgment}

K.M. would like to thank the Japan Society for the
Promotion of Science for financial support.
This work was supported by the Grant-in-Aid for Scientific Research
 on Innovative Areas (No.21111006 [K.N.]) and Scientific Research (A) (No.22244030 [K.N.]).

\appendix 

\section{Plasma dissipation on SCS loops}   \label{App}

A SCS with a current generates magnetic field.
Surrounding plasma flow into a SCS, but the presence of magnetic field prevents the plasma particles to hit the SCS itself.
Thus there is a cylindrical region around SCS where plasma cannot invade~\cite{Vilenkin}.
The radius of this cylindrical region is given by $R_s \sim I/(u\sqrt{\rho_e})$,
where $u$ is the velocity of the SCS and $\rho_e$ the energy density of the electron plasma.
The force acting on the SCS per unit length is estimated as
\begin{equation}
	F_{\rm dis} \sim \begin{cases}
		\rho_e u \nu       & ;~~uR_s < \nu\\
		\rho_e u^2 R_s  & ;~~uR_s > \nu,
	\end{cases}
\end{equation}
where $\nu$ is the viscosity of the plasma.
It is roughly given by $\nu \sim v_e \lambda_e$, where $v_e$ is the average velocity of the electron and $\lambda_e$ is its mean free length.
The velocity $v_e$ varies in time as
\be
v_e\sim
\begin{cases}
1 & ; \ T>m_e \\
\sqrt{\frac{T}{m_e}} & ; \ T_{\rm rec}<T<m_e \\
\frac{T}{T_{\rm rec}}\sqrt{\frac{T_{\rm rec}}{m_e}} &; \ T<T_{\rm rec}
\end{cases},
\ee
where $T_{\rm rec}$ is the temperature at the recombination.
The mean free length is evaluated as $\lambda_e \sim (\sigma_R n_e)^{-1}$
where $n_e$ is the electron number density and $\sigma_R\sim \sigma_T v_e^{-4}$
is the cross section of electron-electron scattering 
with $\sigma_T$ denoting the Thomson scattering cross section.
This frictional force tends to damp the motion of SCSs.
On the other hand, the force caused by the string tension itself is given by
$F_t \sim \mu/l$ with $l$ being the curvature scale of the SCS loop, which is roughly equal to the length of the loop.
If $F_t \gg F_{\rm dis}$, the effect of plasma dissipation is negligible and the dynamics is dominated by the
string tension itself.
If $F_t \ll F_{\rm dis}$, the SCS dynamics will be damped and its motion will be non-relativistic, which might 
affect the evaluation of the GW and EMW emission.
Here we evaluate the ratio of these forces.
For a loop with length $l=\Gamma_{\rm eff}t$, we find
\begin{equation}
	\frac{F_t}{F_{\rm dis}} \sim \frac{G\mu}{u\Gamma_{\rm eff}}\frac{\sigma_T}{Gm_et}v_e^{-5}, \label{FtFd1}
\end{equation}
if $uR_s < \nu$, and
\begin{equation}
	\frac{F_t}{F_{\rm dis}} \sim \frac{G\mu}{u\Gamma_{\rm eff}I}\frac{1}{G\sqrt{\rho_e}t}, \label{FtFd2}
\end{equation}
if $uR_s > \nu$.
Eq. (\ref{FtFd1}) is minimized at $T=m_e$ and the value at that time is
\be
\frac{F_t}{F_{\rm dis}} \sim 10^{19}\times\frac{G\mu}{u\Gamma_{\rm eff}}.
\ee
The minimum value of Eq. (\ref{FtFd2}), which is taken at the matter-radiation equality, is
\be
\frac{F_t}{F_{\rm dis}} \sim 10^{22}\times\frac{G\mu}{u\Gamma_{\rm eff}}\left(\frac{1{\rm GeV}}{I}\right).
\ee
Therefore, the drag force due to plasma dissipation can be neglected compared with the string tension and the assumption that SCS motions are relativistic $(u\sim 1)$ is justified.

However, the dissipation due to the plasma interaction still may not be negligible  as a energy loss mechanism of a loop.
The energy loss rate of a SCS loop due to the plasma interaction is estimated as
\begin{equation}
	P_{\rm dis} \sim F_{\rm dis} l .
\end{equation}
This should be compared with the EMW and GW emission rate, $P_{\rm EW}$ and $P_{\rm GW}$. We find
\begin{equation}
	\frac{P_{\rm GW}}{P_{\rm dis}} \sim \frac{F_t}{F_{\rm dis}}\Gamma_{\rm GW} G\mu,
\end{equation}
and
\begin{equation}
	\frac{P_{\rm EM}}{P_{\rm dis}} \sim \frac{F_t}{F_{\rm dis}}\Gamma_{\rm EM} \frac{I}{\sqrt{\mu}}.
\end{equation}
Although these ratios are larger than 1 for most parameter sets we are interested in, they can be smaller for very small $G\mu$ or $I$.
In fact, constraint from the CMB anisotropy, CMB spectral distortion and radio wave observation, which include the very small values of $G\mu$ and $I$, are affected by the dissipation effect.

{}

\end{document}